\documentclass[fleqn,usenatbib]{mnras}

\usepackage{newtxtext,newtxmath}
\usepackage{float}

\usepackage[T1]{fontenc}

\DeclareRobustCommand{\VAN}[3]{#2}
\let\VANthebibliography\thebibliography
\def\thebibliography{\DeclareRobustCommand{\VAN}[3]{##3}\VANthebibliography}

\usepackage{graphicx}	
\usepackage{amsmath}	

\newcommand{\Msun}{\mathrm{M_{\sun}}}

\newcommand*{\vcenteredhbox}[1]{\begin{tabular}{@{}c@{}}#1\end{tabular}}

\title[XRBs in NGC 362]{Chandra and HST studies of the X-ray sources in the Globular Cluster NGC 362}

\author[Kumawat et al.]{Gourav Kumawat,$^{1,2}$\thanks{E-mail: gourav19@iiserb.ac.in}
Craig O. Heinke,$^{2}$\thanks{E-mail: heinke@ualberta.ca}
Haldan N. Cohn,$^3$
Phyllis M. Lugger$^3$
\\
$^{1}$Department of Physics, Indian Institute of Science Education and Research, Bhopal 462066, India\\
$^{2}$ Department of Physics, University of Alberta, Edmonton, AB T6G 2G7, Canada\\
$^3$ Department of Astronomy, Indiana University, 727 E. Third St., Bloomington, IN 47405, USA
}

\date{Accepted XXX. Received YYY; in original form ZZZ}

\pubyear{2023}

\begin{document}
\label{firstpage}
\pagerange{\pageref{firstpage}--\pageref{lastpage}}
\maketitle

\begin{abstract} 
We analyse a {\it Chandra} observation of the rich globular cluster NGC 362, finding 33 X-ray sources within 1' (1.2 half-mass radii) of the cluster center. 
Spectral analysis of the brightest source (X1) shows blackbody-like emission, indicating it is likely a quiescent low-mass X-ray binary; we find a possible counterpart that falls in the sub-subgiant region. We use {\it HST} UV Globular Cluster Survey (HUGS) photometry to identify 15 potential optical/UV counterparts to these X-ray sources, including two background AGN. 
We identify no likely CVs, probably due to crowding in optical filters in the core, 
though we predict of order 8 CVs among the detected X-ray sources.
We identify three other sub-subgiants and two red straggler counterparts, which are likely powered by coronal activity, along with five other potential coronally active binary counterparts to three X-ray sources. Finally, we note two unusual counterpart candidates that lie to the red of the red giant branch in $V_{606}-I_{814}$, and shift well to the blue of the red giant branch in ultraviolet colour-magnitude diagrams. These systems seem to contain a red giant with a distorted evolutionary history, plus a bright blue light source, either a blue straggler star (an Algol-like system) or an accreting white dwarf (a long-period CV, or a symbiotic star). 

\end{abstract}




\section{INTRODUCTION}
Globular clusters (GCs) are stellar systems characterized by high densities, leading to frequent gravitational interactions between stars. These interactions give rise to various dynamical processes, including collisions and mass transfers, which can result in the formation of exotic binary star systems such as millisecond pulsars (MSPs), cataclysmic variables (CVs), and low-mass X-ray binaries (LMXBs), among others.  Early observations of bright LMXBs (in Galactic globular clusters, generally neutron stars accreting from low-mass companion stars via an accretion disc) showed that they were of order 100 times more common in globular clusters per unit mass than the rest of the Galaxy \citep{Clark75}, and indeed the clusters with the highest predicted rates of stellar interaction are the most likely to contain LMXBs \citep{Verbunt87}. This can be attributed to the tendency of LMXB progenitor binaries to become unbound gravitationally when the more massive star goes supernova; this makes LMXBs rare in the field, but neutron stars can gain a new partner in globular clusters. 

The study of faint X-ray binaries with $L_X$ < 10$^{34.5}$ erg/s in GCs was pioneered by \cite{1983ApJ...267L..83H}. The advent of the \textit{Chandra} X-Ray Observatory and the Hubble Space Telescope (\textit{\textit{HST}}) with their enhanced resolution and improved sensitivity made it feasible to investigate faint X-ray binaries within the dense stellar environments of GCs. Among these faint X-ray binaries, the brightest are classified as quiescent LMXBs (qLMXBs), which are LMXBs in which accretion onto the neutron star is low to nonexistent. 
X-ray emission from qLMXBs can occur through thermal emission from the neutron star surface or continued low-level accretion onto the neutron star. 
Many qLMXB X-ray spectra are dominated by  bright, soft blackbody-like X-ray emission, enabling clear identification in globular clusters \citep{Rutledge02a,Heinke03d,Guillot09}.
These systems are primarily formed in GCs through stellar encounters, and their numbers are correlated with the total encounter rate of the cluster \citep{Heinke03d,Pooley06}. 
While their optical counterparts appear blue, the accretion disk is generally faint, as their mass transfer rates are low \citep[e.g.][]{Haggard04}, leading to high X-ray/optical flux ratios \citep{Verbunt08}.

MSPs are believed to be the descendants of LMXBs \citep{Alpar82}, as the neutron star magnetic field is reduced and the neutron star spin increased during the recycling process. Their X-ray emission may include hard nonthermal synchrotron pulsed emission from their magnetosphere \citep{Saito97}, hard nonthermal shock emission from a collision between the pulsar wind and a stellar wind \citep{Wadiasingh17}. However, X-rays from most MSPs are dominated by thermal blackbody-like emission from the heated polar caps of the neutron star, such that they are typically soft X-ray sources with $L_X\sim10^{30-31}$ erg/s \citep{Zavlin02,Bogdanov06}. MSP numbers are also correlated with the total encounter rate of the cluster \citep{Bagchi11,Bahramian13}.  
Their optical counterparts may be helium white dwarfs \citep[e.g.][]{Edmonds01,Chen23}, or nondegenerate stars, so-called 'spiders', which may show unusual evolution and heating \citep{Ferraro01,Cadelano15}.

Cataclysmic variables (CVs) are white dwarfs (WDs) accreting from a main-sequence or subgiant companion star, via Roche Lobe Overflow (RLOF). 
Their X-rays are produced by thermal hot plasma, leading to relatively hard X-ray spectra and X-ray luminosities of typically $L_X<10^{32}$ erg/s \citep[e.g.][]{Heinke05}.
While CVs are formed in the densest globular clusters, their progenitors are also destroyed in many globular clusters, and the effects of dynamics on their population is the subject of current research  \citep{Cool13,Cheng18,Belloni19,Heinke20}.
White dwarfs may accrete from evolved (red giant or subgiant) stars via either Roche-lobe overflow or from a stellar wind. In the latter case, they are known as symbiotic stars \citep{2013A&A...559A...6L}. (If a neutron star or black hole accretes from a red giant wind, this is termed a symbiotic X-ray binary.)
No confirmed symbiotic stars have been identified in globular clusters \citep{Henleywillis18,Bahramian20}, which may be explained by the destruction of wide binaries in clusters \citep{Belloni20}.

Active binaries (ABs) consist of main-sequence or (sub)giant stars in a binary system that emit X-rays due to coronal activity on one or both stars \citep{Gudel04}. 
Their numbers are not increased by dynamical processes in clusters; instead, they are likely decreased by the destruction of wide binaries in all clusters \citep{Verbunt00,Cheng18,Heinke20}. To first order, their numbers appear to scale with the total mass of the cluster \citep{Bassa04,Bassa08}.

Active binaries may be identified in several unusual places in colour-magnitude diagrams (CMDs); above the main sequence, below the subgiant branch, or to the red of the giant branch. Binary stars will naturally appear above the main sequence, by up to 0.75 magnitudes (or a factor of 2). 
Sub-subgiants (SSGs) are redder than the main-sequence, but fainter than subgiant stars, while red straggler stars (RSS) lie to the red of the red giant branch \citep{Geller17}. 
Their unusual CMD locations may result from rapid rotation in subgiants or giants due to tidal synchronization in a close binary system. The strong magnetic fields generated in these stars inhibit convection, leading to the formation of large starspots, radius inflation, and lower-than-expected average surface temperatures and luminosities \citep{2022ApJ...927..222L}. The emission of X-rays from SSGs is a consequence of these strong magnetic fields.

The globular cluster NGC 362 is 
located 8.8 kpc away \citep{Baumgardt21}, $\sim$1 degree north of the northern end of the Small Magellanic Cloud's (SMC) bar.
The cluster experiences a reddening of approximately 0.05 mag, and its half-mass radius is determined to be 0.82 arcmin \citep[][2010 revision]{Harris96}. 
Its metallicity is \big[Fe/H\big]=-1.3, and its age is estimated at 11 Gyr \citep{2021MNRAS.505.5978V}. 
It has a relatively high mass ($2.5\times10^5$ $\Msun$, \citealt{2021MNRAS.505.5978V}), may be core-collapsed \citet[][2010 revision]{Harris96}, and has one of the highest estimated stellar encounter rates among Galactic globular clusters, $0.74\pm0.13$ the rate for 47 Tucanae \citep{Bahramian13}.
A preliminary report of \textit{Chandra} and \textit{\textit{HST}} analyses of NGC 362 X-ray binaries was made by \citet{Margon10}, who suggested that one X-ray source is a qLMXB.  
Here we analyse the archival \textit{Chandra} exposure and use photometric catalogues produced by the \textit{HST} UV Globular Cluster Survey \citep{Piotto15,Nardiello18} to identify probable optical/UV counterparts.

\section{HST CATALOGUES USED}

NGC 362 was observed by the Wide Field Camera 3 (WFC3; GO-12605) and Advanced Camera for Surveys (ACS; GO-10775) on board the \textit{HST}. GO-12605 (PI: Piotto) contains exposures in two UV filters, F275W (hereafter $UV_{275}$) and F336W ($U_{336}$), along with an exposure in F438W ($B_{438}$) taken in 2012, while GO-10775 (PI: Sarajedini) is comprised of exposures in F606W ($V_{606}$) and F814W ($I_{814}$) taken in 2006 (see Table~\ref{tab:HSTobservations}). For all the filters, we retrieved photometric and astrometric catalogues and astrometrised stacked images from the \textit{\textit{HST}} UV Globular Cluster Survey ("HUGS", \cite{2015AJ....149...91P}).

\begin{table}
\centering
 \caption{\textit{\textit{HST}} Observations of NGC 362}
\begin{tabular}{ccccc} 
 \hline
    Telescope/ & Date of & Proposal & Filter & Exposure \\ 
    instrument & Observation       & ID & & time (s) \\ 
    \hline
\textit{\textit{HST}}/WFC3 & 2012 Sep 09 &  GO-12605 & F275W & 1557 \\
\textit{\textit{HST}}/WFC3 & 2012 Sep 09 &  GO-12605 & F336W & 700 \\
\textit{\textit{HST}}/WFC3 & 2012 Sep 09 &  GO-12605 & F438W & 108 \\ 
\textit{\textit{HST}}/ACS & 2006 Jun 02 &  GO-10775 & F606W & 610 \\
\textit{\textit{HST}}/ACS & 2006 Jun 02 & GO-10775 & F814W & 690 \\
\hline
\end{tabular}
 \label{tab:HSTobservations}
\end{table}

 The HUGS data reduction procedure is extensively described in \cite{2008AJ....135.2055A} and \cite{2017ApJ...842....6B}; we summarise it briefly here. 
In the initial "first-pass" photometry stage, these authors utilized the \texttt{FORTRAN} program \texttt{HST1pass} to generate perturbed PSFs for each image. Bright, unsaturated, and isolated stars were selected, and their fluxes and positions were measured using the library PSFs. A model of each star was then subtracted from the real star. Preliminary catalogues were extracted using the \texttt{HST1pass} program on these PSF arrays. 
In the subsequent "multi-pass" photometry stage, the authors employed the images, PSF arrays, and transformations obtained during the "first-pass" stage to simultaneously detect and measure stars in all individual exposures, 
using the \texttt{FORTRAN} software \texttt{kitchen\_sync2} (KS2; \citealt{2017ApJ...842....6B}). 
To calibrate the photometry output from KS2, the authors compared aperture photometry on \texttt{\_drc} images (normalized to an exposure time of 1 s) with their PSF-based photometry. 
The KS2 software generated astrometric and photometric catalogues of stars using three different methods. Method~1 works best for bright stars generating star-like profiles in single exposures, Method~2 works well for faint stars and crowded environments, and Method~3 works well for highly crowded environments. In this paper, we used Method~3 photometry, since this is a crowded cluster.

The astrometric solution was based on the \textit{Gaia} Data Release 1 catalogue \citep{2016A&A...595A...2G} as the reference frame. To determine the membership probabilities of the target stars, the authors employed the local-sample method \citep{2009A&A...493..959B}. For each target star, membership probability was estimated by considering a sub-sample of 500 reference stars from the catalogue. The selection of these reference stars was based on their proper motion error (typically  $\pm$ 0.25 mas yr$^{-1}$) and a magnitude similar to that of the target star. The catalogue indicates whether a membership probability calculation is not possible (value -1), or gives a probability estimate for belonging to the cluster. 

\section{X-ray Observations and Data Reduction}

The X-ray data utilized in this study comprised two \textit{Chandra} X-ray Observatory observations of NGC 362 in 2004, totaling an exposure time of 78.82 ks (see Table \ref{tab:chandraobservations}). For both observations, the core of NGC 362 was positioned on the back-illuminated ACIS-S3 chip and configured in FAINT mode.

\begin{table}
\centering
 \caption{\textit{Chandra} Observations of NGC 362}
\begin{tabular}{cccc} 
 \hline
    Telescope/ & Date of & Observation & Exposure \\ 
    instrument & Observation       & ID & time (ks) \\ 
    \hline
\textit{Chandra}/ACIS-S  &2004 Jan 30 &4529  &65.85\\
\textit{Chandra}/ACIS-S  &2004 Jan 27 &5299  &12.97\\
\hline
\end{tabular}
 \label{tab:chandraobservations}
\end{table}

We performed data reduction and analysis using CIAO\footnote{Chandra Interactive Analysis of Observations; available at \href{https://cxc.cfa.harvard.edu/ciao/}{https://cxc.cfa.harvard.edu/ciao/}} (version 4.15.1, CALDB 4.10.4) provided by the \textit{Chandra} X-ray Center (CXC). The data were reprocessed using the \texttt{chandra\_repro} script, which generated new level 2 event files for each observation incorporating the latest calibration updates and bad pixel files. We applied fine astrometric shifts to the shorter observation (ID 5299) using the \texttt{reproject\_aspect} script, aligning it with the longer observation (ID 4299). Finally, the event files from both observations were merged using the \texttt{reproject\_obs} script.

\subsection{Source Detection}
We utilized the CIAO package \texttt{wavdetect} for the detection of X-ray sources. Two regions were defined around the cluster's center: a larger region with an 8 arcminute radius and a smaller region with a radius of 1 arcminute ($\approx 1.2 \times$ the cluster's half-mass radius). The data in the larger region were filtered to the energy range of 0.5-8 keV. The wavelet scales were specified as a series from 1 to 8, increasing by a factor of $\sqrt{2}$, and a significance threshold for source detection was set at $10^{-6}$ (false sources per pixel). 

To refine the astrometry of the \textit{Chandra} dataset, we cross-matched the detected X-ray sources with the \textit{Gaia} DR3 dataset \citep{2023A&A...674A...1G} within a search radius of 1 arcsec. The offset between \textit{Gaia} and \textit{Chandra} astrometry was found to be significantly smaller than the position uncertainties of the \textit{Chandra} sources (see section \ref{sec:4}). Therefore, this method did not yield further improvements in astrometric accuracy. We note that the proper motion offset between Gaia DR1 and DR3 positions is small compared to our X-ray positional uncertainties.

\begin{figure*}
\centering
\includegraphics[scale=0.25]{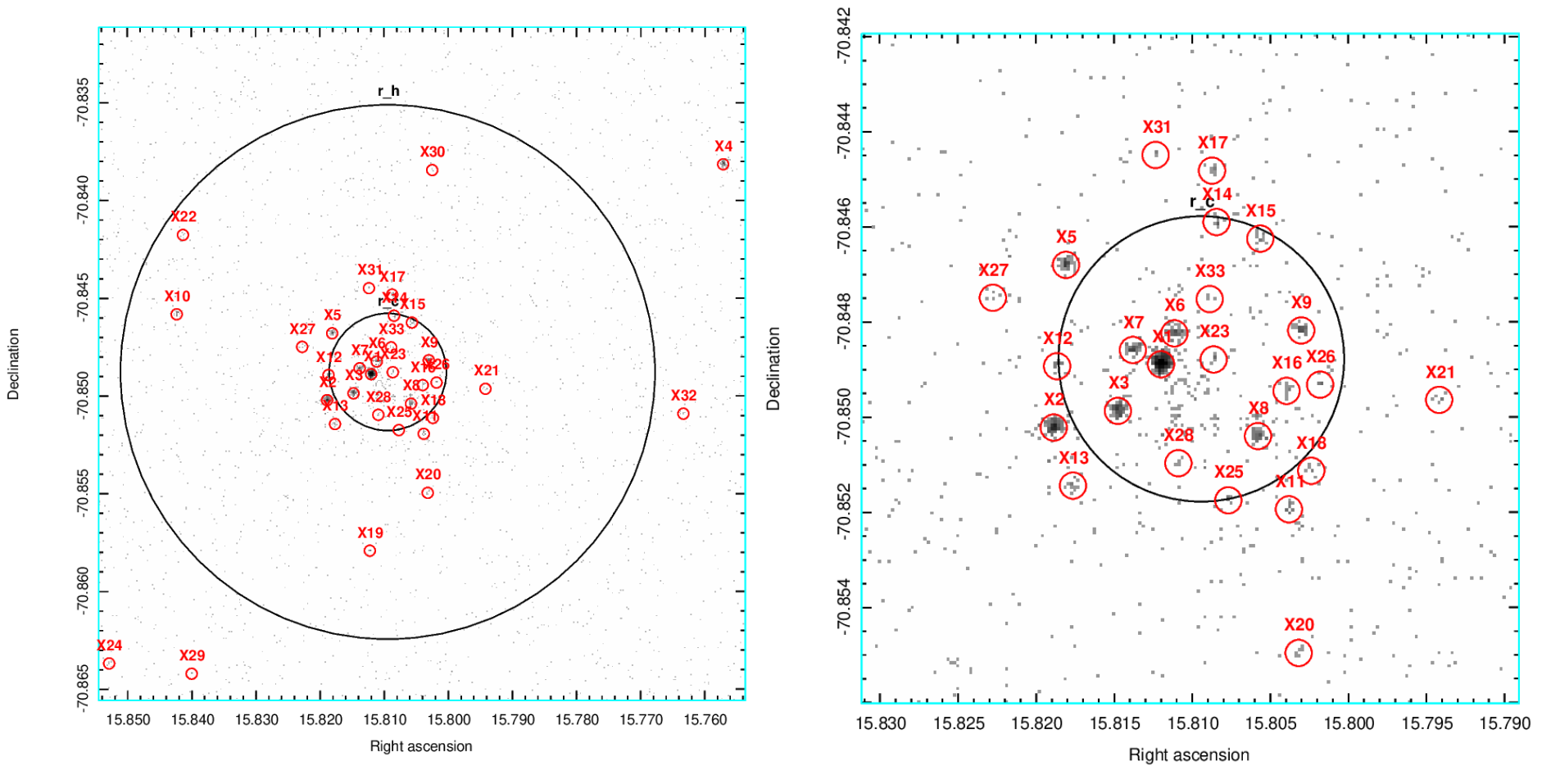}
\caption{Merged Chandra ACIS-S 0.5–8 keV image of NGC 362. The 33 detected X-ray sources are marked with red circles of 1 arcsec radius and labeled with their source IDs. The inner black circle represents the core radius, r\_c (0.18 arcmin), and the larger black circle represents the half-mass radius, r\_h (0.82 arcmin), of the globular cluster NGC 362.  The left panel shows all detected Chandra sources, while the right panel shows the detected sources in and around the core of NGC 362. Additional diffuse emission (likely faint point sources) is also visible in the core.}
\label{fig:clusterxrayimg}
\end{figure*}

Within the smaller region, the data were filtered into three distinct energy ranges: 0.5-2 keV, 2-8 keV, and 0.5-8 keV. For each energy range, the same wavelet scales as in the case of the larger region were employed. The detection signal threshold was set to $10^{-5}$. The unique X-ray source detections from each energy range were subsequently combined, resulting in a master file containing 33 X-ray sources (see Table \ref{tab:xraysources}). The X-ray sources are shown in Figure \ref{fig:clusterxrayimg}.

\begin{table}
\centering
 \caption{Properties of \textit{Chandra} sources. The right ascension (R.A.), declination (Decl.), and net counts in 0.5-8 keV are provided by the \texttt{wavdetect} algorithm. The RA and Decl. are corrected for the offset between {\it Chandra} and {\it HST} positions of three sub-subgiants. The unabsorbed luminosities are derived from power-law spectral fitting, as discussed in Section \ref{sec:3.3}. For sources with net counts $\le$ 10, the luminosities are extrapolated using a linear relation between net counts and luminosity. The positional uncertainty is given in arcseconds (see text).}
\begin{tabular}{ccccccc} 
 \hline
    Source & R.A. & Decl. & Net & $L_{X(0.5-10)}$ & Uncert. \\
    ID & (J2000.0) & (J2000.0) & Counts & ($10^{32}$ erg/s) & (") \\
    \hline
X1	&	1:03:14.880	&	-70:50:56.00	&	598	$\pm$	25	&	4.56	&	0.16	\\
X2	&	1:03:16.529	&	-70:51:00.79	&	125	$\pm$	12	&	1.60	&	0.20	\\
X3	&	1:03:15.542	&	-70:50:59.53	&	81	$\pm$	9	&	1.49	&	0.22	\\
X4	&	1:03:01.716	&	-70:50:17.34	&	71	$\pm$	9	&	0.88	&	0.29	\\
X5	&	1:03:16.339	&	-70:50:48.48	&	55	$\pm$	8	&	0.91	&	0.26	\\
X6	&	1:03:14.674	&	-70:50:53.66	&	41	$\pm$	7	&	0.68	&	0.28	\\
X7	&	1:03:15.312	&	-70:50:54.92	&	40	$\pm$	7	&	0.73	&	0.29	\\
X8	&	1:03:13.390	&	-70:51:01.44	&	34	$\pm$	6	&	0.37	&	0.31	\\
X9	&	1:03:12.725	&	-70:50:53.45	&	32	$\pm$	6	&	0.33	&	0.32	\\
X10	&	1:03:22.154	&	-70:50:44.95	&	17	$\pm$	4	&	0.27	&	0.48	\\
X11	&	1:03:12.914	&	-70:51:06.98	&	13	$\pm$	4	&	0.12	&	0.48	\\
X12	&	1:03:16.476	&	-70:50:56.15	&	11	$\pm$	3	&	0.13	&	0.51	\\
X13	&	1:03:16.234	&	-70:51:05.18	&	11	$\pm$	4	&	0.51	&	0.52	\\
X14	&	1:03:14.026	&	-70:50:45.24	&	11	$\pm$	3	&	0.12	&	0.52	\\
X15	&	1:03:13.354	&	-70:50:46.50	&	10	$\pm$	3	&	0.10	&	0.54	\\
X16	&	1:03:12.950	&	-70:50:58.02	&	10	$\pm$	3	&	0.10	&	0.55	\\
X17	&	1:03:14.095	&	-70:50:41.35	&	7	$\pm$	3	&	0.07	&	0.63	\\
X18	&	1:03:12.571	&	-70:51:04.07	&	6	$\pm$	3	&	0.06	&	0.68	\\
X19	&	1:03:14.933	&	-70:51:28.51	&	6	$\pm$	2	&	0.06	&	0.76	\\
X20	&	1:03:12.766	&	-70:51:17.86	&	6	$\pm$	2	&	0.06	&	0.74	\\
X21	&	1:03:10.606	&	-70:50:58.70	&	6	$\pm$	2	&	0.06	&	0.73	\\
X22	&	1:03:21.917	&	-70:50:30.34	&	5	$\pm$	2	&	0.05	&	0.84	\\
X23	&	1:03:14.071	&	-70:50:55.64	&	5	$\pm$	2	&	0.05	&	0.74	\\
X24	&	1:03:24.694	&	-70:51:49.25	&	5	$\pm$	2	&	0.05	&	1.02	\\
X25	&	1:03:13.846	&	-70:51:06.30	&	5	$\pm$	2	&	0.05	&	0.78	\\
X26	&	1:03:12.434	&	-70:50:57.55	&	5	$\pm$	2	&	0.05	&	0.77	\\
X27	&	1:03:17.462	&	-70:50:50.96	&	3	$\pm$	2	&	0.03	&	0.93	\\
X28	&	1:03:14.614	&	-70:51:03.49	&	3	$\pm$	2	&	0.03	&	0.91	\\
X29	&	1:03:21.605	&	-70:51:51.16	&	3	$\pm$	2	&	0.03	&	1.23	\\
X30	&	1:03:12.595	&	-70:50:18.42	&	3	$\pm$	2	&	0.03	&	1.10	\\
X31	&	1:03:14.962	&	-70:50:40.16	&	3	$\pm$	2	&	0.03	&	1.00	\\
X32	&	1:03:03.199	&	-70:51:03.28	&	3	$\pm$	2	&	0.03	&	1.22	\\
X33	&	1:03:14.131	&	-70:50:51.07	&	3	$\pm$	2	&	0.03	&	1.00	\\
\hline
\end{tabular}
 \label{tab:xraysources}
\end{table}

\subsection{Spectral Extraction}

We utilized the \texttt{specextract} script to extract spectra for the detected Chandra sources with "Net Counts" > 10 (see Table \ref{tab:xraysources}) for both observations 4529 and 5299 (see Table \ref{tab:chandraobservations}). The \texttt{specextract} script is designed for creating source and optional background spectra in Chandra imaging mode.

For the background region, we selected a source-free area with a radius of 25 arcsec near the cluster. As for the source regions, they were chosen to have a radius of 1 arcsec for each source. To combine the spectra from the two observations, we utilized the \texttt{combine\_spectra} script. This script performs the summation of multiple PHA spectra and combines the associated background PHA spectra, as well as the source and background ARF and RMF response files.

\subsection{Spectral Fitting}\label{sec:3.3}
We utilized CIAO's modelling and fitting application, \texttt{SHERPA}\footnote{ \href{https://cxc.cfa.harvard.edu/sherpa/}{https://cxc.cfa.harvard.edu/sherpa/}} for all spectral fits. To account for the limited number of 
counts  
in the X-ray spectra, we employed the \texttt{WSTAT}\footnote{\href{https://heasarc.gsfc.nasa.gov/xanadu/xspec/manual/XSappendixStatistics.html}{https://heasarc.gsfc.nasa.gov/xanadu/xspec/manual/XSappendixStatistics.html}} statistic, which is a Poisson log-likelihood function that includes a Poisson background component. 
Considering the limited counts, 
we grouped the spectral data to ensure that each bin contains at least one photon (\cite{2009ApJ...693..822H}). 

Initially, we performed fits on all spectra using a combination of a power law with pegged normalization (\texttt{xspegpwrlw}) and XSpec photoelectric absorption (\texttt{xsphabs}) as the source model. The hydrogen column density ($N_H$) was set to $4.4\times10^{20}$ cm$^{-2}$, calculated by multiplying the optical extinction \textit{E(B-V)}=0.05 \citep{2010arXiv1012.3224H} by a factor of 3.1 to convert to $A_V$, and by a conversion of $2.81\times10^{21}$ cm$^{-2}$ to convert to $N_H$ \citep{2015MNRAS.452.3475B}. The 
energy range for 
the normalization was chosen to be 0.5 to 10 keV. 
We set the elemental abundances used by XSpec models to \texttt{wilm} \citep{2000ApJ...542..914W}. This initial model yielded satisfactory results for all sources, except for source X1 (see Table \ref{tab:powerlawresults},  and below). We extrapolated a relation between the net counts and power-law $L_X$ to infer estimated $L_X$ for the fainter sources as well, in Table~\ref{tab:xraysources}. We do not provide formal $L_X$ error estimates in Table~\ref{tab:xraysources}; the dominant $L_X$ error should be small-number count statistics, but uncertainty in spectral shape is also substantial.

\begin{table}
\centering
 \caption{Spectral fit parameters of the \textit{Chandra} sources with \texttt{NET\_COUNTS} > 10. The model used is  a power law with pegged normalization (\texttt{xspegpwrlw}) and photoelectric absorption (\texttt{xsphabs}) fixed to $N_H=4.4\times10^{20}$ cm$^{-2}$, and $L_X$ values reported are corrected for absorption.}
\begin{tabular}{ccc} 
 \hline
    Source ID & Photon Index & $L_{X(0.5-10)}$ (10$^{32}$ erg/s) \\ 
    \hline
X1  & 3.2 $\pm$ 0.1 & 4.56 $\pm$ 0.18 \\
X2  & 1.7 $\pm$ 0.1 & 1.60  $\pm$ 0.21 \\
X3  & 1.3 $\pm$ 0.2 & 1.49 $\pm$ 0.25 \\
X4  & 1.7 $\pm$ 0.2 & 0.88 $\pm$ 0.15 \\
X5  & 1.4 $\pm$ 0.2 & 0.91 $\pm$ 0.18 \\
X6  & 1.5 $\pm$ 0.2 & 0.68 $\pm$ 0.16 \\
X7  & 1.4 $\pm$ 0.2 & 0.73 $\pm$ 0.17 \\
X8  & 1.8 $\pm$ 0.2 & 0.37 $\pm$ 0.10  \\
X9  & 2.0   $\pm$ 0.2 & 0.33 $\pm$ 0.09 \\
X10 & 2.0   $\pm$ 0.3 & 0.27 $\pm$ 0.07 \\
X11 & 2.5 $\pm$ 0.3 & 0.12 $\pm$ 0.04 \\
X12 & 2.5 $\pm$ 0.2 & 0.13 $\pm$ 0.04 \\
X13 & 0.7 $\pm$ 0.3 & 0.51 $\pm$ 0.20  \\
X14 & 1.9 $\pm$ 0.3 & 0.12 $\pm$ 0.06 \\   
\hline
\end{tabular}
 \label{tab:powerlawresults}
\end{table}

Source X1 showed an unusually steep power-law spectrum, with a photon index of 3.2 $\pm$ 0.1, which suggests that it might be better fit by a blackbody-like neutron star atmosphere model.
To address this, 
we used 
an XSpec photoelectric absorption model (\texttt{xsphabs}) with a low-magnetic-field NS hydrogen atmosphere model 
(\texttt{xsnsatmos}, \citealt{2006ApJ...644.1090H}). In this fit, the neutron star gravitational mass (\texttt{M\_ns}) was fixed at 1.4 M$_{\odot}$, while the neutron star radius (\texttt{R\_ns}) was fixed at 11.5 km. The distance to the neutron star (\texttt{dist}) was set to 8.8 kpc \citep{Baumgardt21}. This modified model successfully fit Source 1, resulting in a fraction of the neutron star surface emitting (\texttt{norm}) of 1.30$\pm$0.26 and a logarithm of the unredshifted effective temperature (\texttt{Log\_Teff}) of 6.01$\pm$0.02 K, and $L_X$(0.5-10 keV)$=4.2\times10^{32}$ erg/s.
We interpret the close match of the neutron star normalization with the model value of 1.0 as evidence for consistency with the model. We do not attempt to constrain the neutron star radius independently with this data.

During our analysis, we encountered fainter sources that exhibited soft spectra with steep power-law indices ranging between 2 and 2.5 (see Table \ref{tab:powerlawresults}). However, when attempting to fit these sources using the \texttt{xsnsatmos} model, we observed that the resulting normalization parameter (\texttt{norm}) 
was much smaller (0.0004 to 0.001) than the expected value of $\sim$1 (indicating emission from the full surface). 
This does not seem to match observations of quiescent LMXB spectra; though they may be consistent with hot polar spots on millisecond pulsars, typically radio millisecond pulsars showing thermal spectra are not observed to be this luminous \citep{Zhao22}.

\section{COUNTERPART IDENTIFICATION}\label{sec:4}
We performed a search for the optical counterparts of the \textit{Chandra} X-ray sources by crossmatching them with the \textit{HST} {\it HUGS} catalogue using TOPCAT \citep{2005ASPC..347...29T}. We initially used a search radius of 0.6 arcsec for crossmatching.  We identified three promising sub-subgiant candidate counterparts (X15, X16, and X20, see Table \ref{tab:interesting_stars}) with $<$0.2 arcsecond separations. Sub-sub-giants are often identified with X-ray sources (see details below and \citealt{Geller17}), so we identified these as promising counterparts.

To refine the \textit{Chandra} astrometry, we 
shifted the Chandra positions to match the \textit{HST} positions for these three sub-subgiants. The average R.A. offset between the \textit{Chandra} and \textit{HST} catalogues was 0.04", with a standard deviation of 0.13". The average Decl. offset was -0.11", with a standard deviation of 0.02", producing an uncertainty in the optical/X-ray alignment of 0.13".  

\begin{table*}
\centering
 \caption{List of interesting \textit{HST} counterparts to the \textit{Chandra} source detections. The $\Delta$RA and $\Delta$Decl. columns represent the differences between the positions of the \textit{HST} source and the associated \textit{Chandra} source in arcsec. Photometry is from {\it HUGS} ('--' if not detected). The PROB column indicates the cluster membership probability provided in the HUGS catalogue (-1 indicates the probability could not be determined).}
\begin{tabular}{ccccccccccc} 
 \hline
    Source ID & HUGS ID & $\Delta$RA & $\Delta$Decl. & UV$_{275}$ & U$_{336}$ & B$_{438}$ & V$_{606}$ & I$_{814}$ & PROB  & NOTES \\ 
    \hline
X1   & R0077953 & -0.02 & 0.00     & 19.29 & 18.50    & 18.83  & 18.15    & 17.50    & 75.4 & qLMXB        \\
X4   & R0113456 & -0.06 & 0.01  & 22.20  & 21.87  & 23.81   & 22.40    & 21.63   & 0    & Galaxy    \\
X9   & R0005396 & 0.13   & -0.03 & 18.71 & 17.76    & 17.91  & 17.05 & 16.34 & 98   & RG+B \\
X10  & R0006360 & -0.09 & 0.10   & 18.46 & 17.28    & 17.24   & 16.19 & 15.44 & 96.5 & RG+B \\
X12  & R0005116 & 0.05  & 0.11  & 19.72 & 18.21   & 18.24  & 17.19 & 16.44 & 97.8 & RSS       \\
X15  & R0089306 & -0.02 & 0.00     & 20.38 & 19.09   & 18.95  & 18.02   & 17.19   & 93.3 & SSG       \\
X16  & R0078991 & -0.12 & 0.02  & 20.75  & 19.37  & 19.40   & 18.72   & 17.80 & 99.4 & SSG       \\
X20  & R0002866 & 0.14  & -0.04 & 20.39 & 19.10    & 19.03     & 17.83 & 17.11  & 97.8 & SSG       \\
X21A & R0004837 & 0.22  & 0.10   & 19.32 & 17.83   & 17.73 & 16.82  & 16.06 & 95.5 & RSS?       \\
X21B & R0077004 & -0.20  & 0.65  & -- & 23.98    & 23.16   & 21.97   & 21.00      & 97.9 & AB?        \\
X23A & R0078618 & 0.20  & 0.35  & 19.27 & 18.53 & 18.92  & 18.44   & 17.84   & 95.9 & AB?        \\
X23B & R0078683 & 0.39  & 0.38  & 23.16 & 21.16   & 21.27   & 20.27   & 19.56  & 97.1 & AB?        \\
X23C & R0078681 & 0.50  & 0.18  & 20.90  & 20.14    & 20.53   & 20.03  & 19.36  & 99.2 & AB?        \\
X24  & R0029822 & -0.26 & 0.14  & 22.46  & 22.38    & 22.57  & 21.60    & 20.86   & -1   & Galaxy    \\
X26  & R0078462 & 0.00 & 0.19  & 21.86  & 20.79   & 21.05  & 20.31   & 19.65   & 99.2 & AB       \\
\hline
\end{tabular}
 \label{tab:interesting_stars}
\end{table*}

\begin{figure*}
\centering
\includegraphics[scale=0.8]{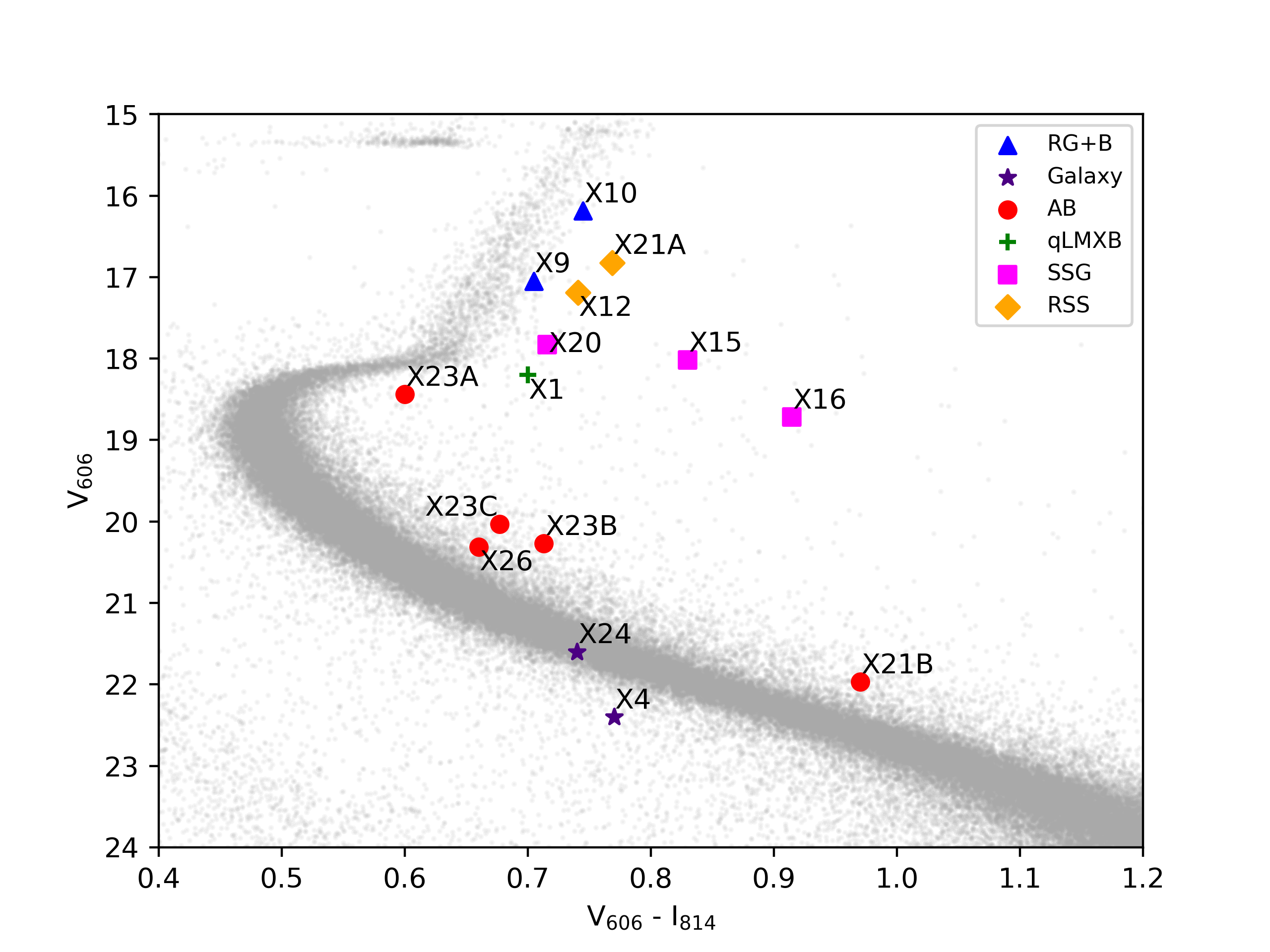}
\caption{ $V_{606}$ vs $V_{606}-I_{814}$ color-magnitude diagram (CMD) from the \textit{HST} ACS {\it HUGS} data on NGC 362. The coloured markers are the positions of the interesting \textit{HST} counterparts to the \textit{Chandra} source detections.}
\label{fig:vi}
\end{figure*}

\begin{figure*}
\centering
\includegraphics[scale=0.8]{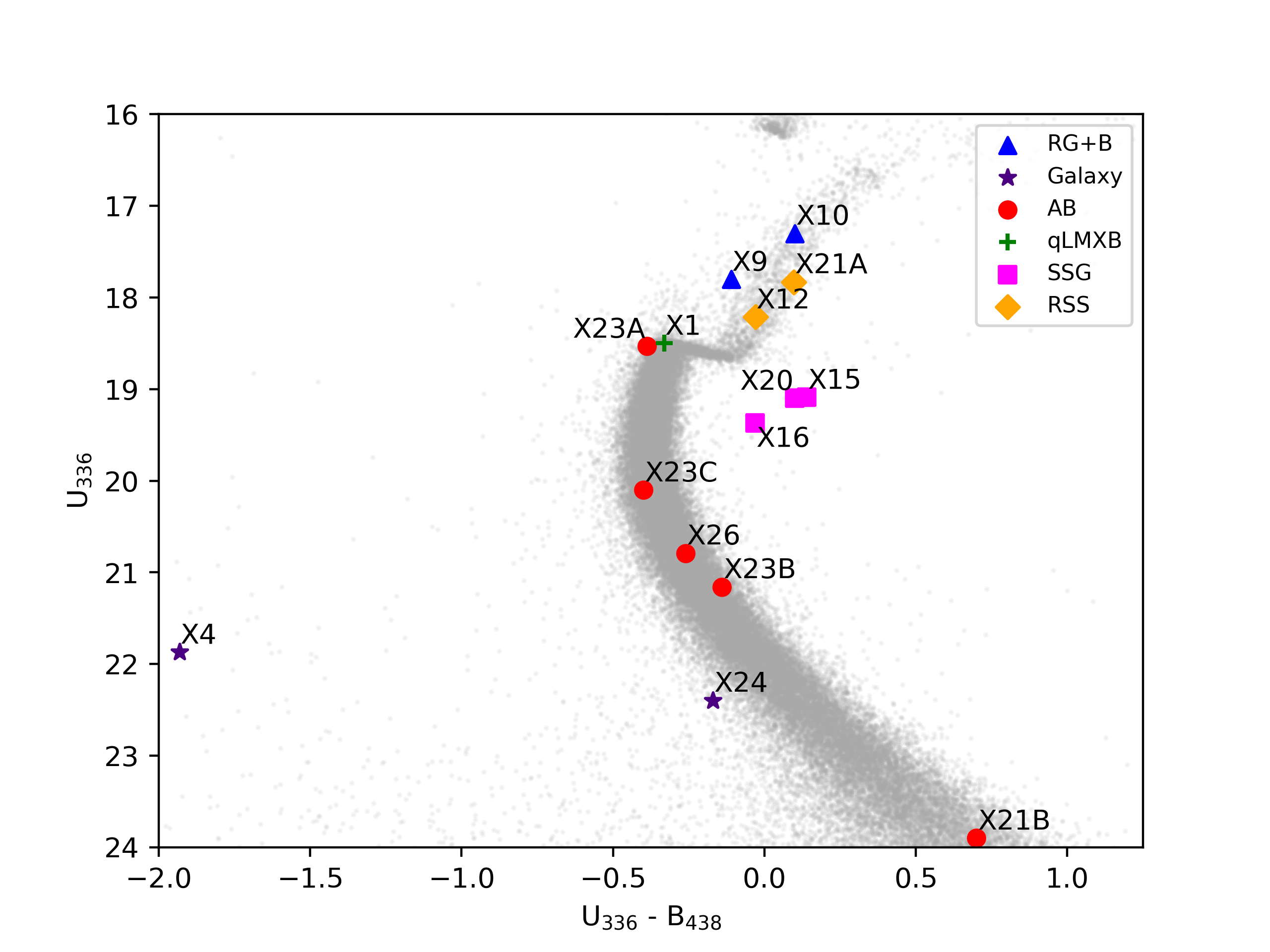}
\caption{$U_{336}$ vs $U_{336}-B_{438}$ CMD of the  \textit{HST} WFC3  {\it HUGS} data on NGC 362. The coloured markers are the positions of the interesting \textit{HST} counterparts to the \textit{Chandra} source detections.}
\label{fig:ub}
\end{figure*}

\begin{figure*}
\centering
\includegraphics[scale=0.8]{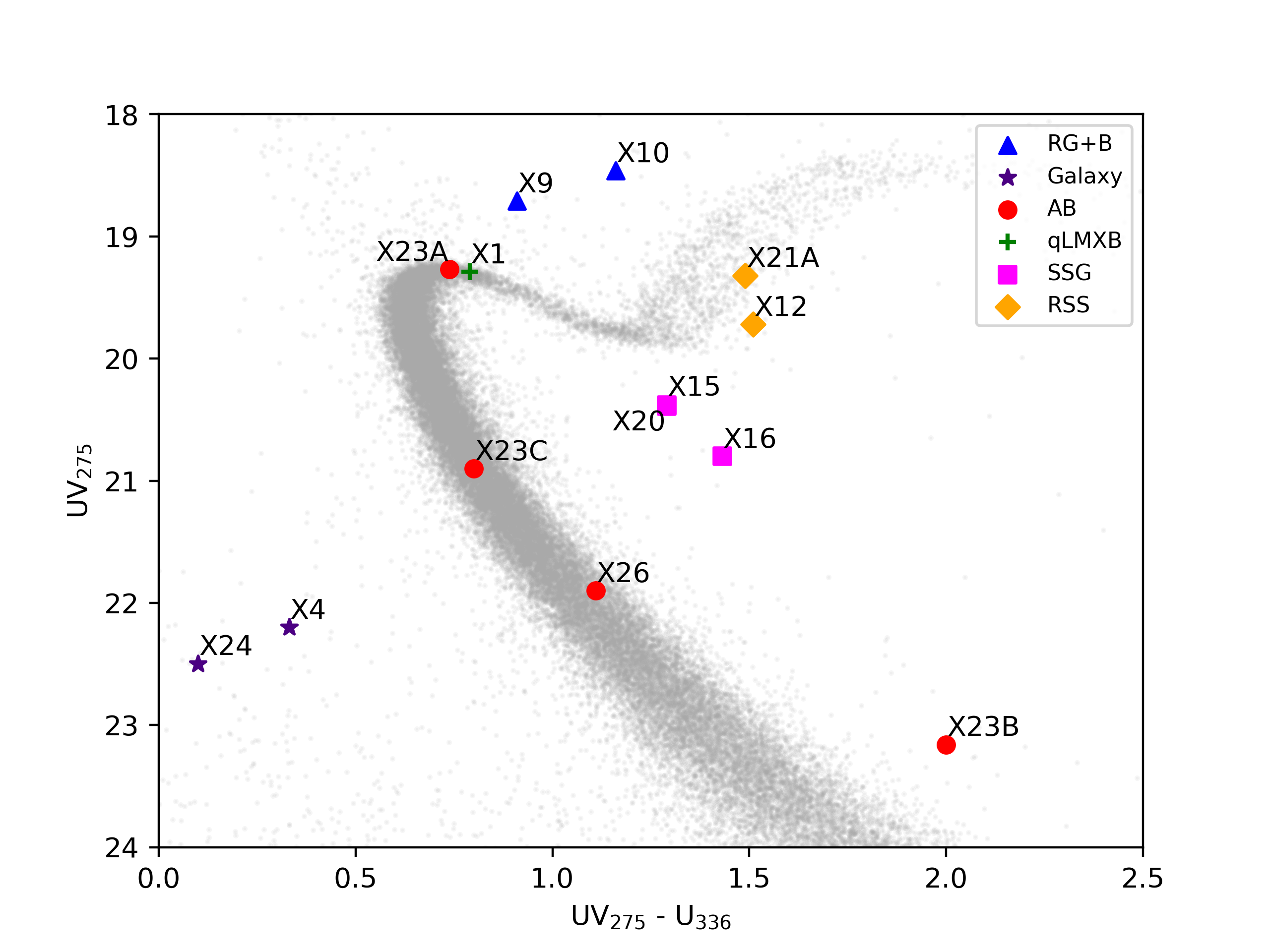}
\caption{$UV_{275}$ vs $UV_{275}-U_{336}$ CMD of the  \textit{HST} WFC3 {\it HUGS} observations of NGC 362. The coloured markers are the positions of the interesting \textit{HST} counterparts to the \textit{Chandra} source detections.}
\label{fig:uvu}
\end{figure*}

Next, we searched for interesting \textit{HST} sources located within the positional uncertainty regions of the \textit{Chandra} X-ray detections shown in  
Table \ref{tab:xraysources}.
The selection of these sources was based on their positions in the {\it HUGS} color-magnitude diagrams (see Figures \ref{fig:vi}, \ref{fig:ub}, \& \ref{fig:uvu}). 

We calculated the {\it Chandra} 95\% confidence positional uncertainties using equation \ref{eqn:PU}  from \cite{2007ApJS..169..401K}, and summing in quadrature the positional uncertainty for each X-ray source with the uncertainty in the X-ray/optical alignment (0.13"). The positional uncertainty, \textit{PU}, is in arcseconds, and off axis angle, \textit{OAA}, is in arcminutes. Source counts, \textit{C}, are as extracted by \texttt{wavdetect}. The final positional uncertainties are listed in Table 2.

\begin{equation}
    \label{eqn:PU}
    logPU = \begin{cases}
      0.1145\times OAA - 0.4958\times \log C + 0.1932,\\
      0.0000<\log C\leq 2.1393\\
      0.0968\times OAA - 0.2064\times \log C - 0.4260,\\
      2.1393<\log C\leq 3.3000
      \end{cases} 
\end{equation}

\subsection{Interesting Candidates}\label{sec:4.1}
We identified a number of candidate counterparts in unusual CMD positions in the three {\it HUGS} CMDs. We discarded potential counterparts if the star appeared significantly blended with another star in the relevant images. 
We include finding charts for all our candidate counterparts in Appendix A. 
We also constructed a plot (Fig.~\ref{fig:FxFv}) of 0.5-10 keV X-ray luminosity vs. $M_V$ (assuming a distance of 8.8 kpc and $m_V-M_V$ of 14.7). We include on the $L_X$/$M_V$ plot equations 2.2 and 2.3 of \citet{Verbunt08}, which are empirical limits on the $L_X$ of ABs at a given $M_V$ estimated from reported matches in the globular cluster X-ray source population literature, and from ROSAT observations of nearby ABs, respectively. 

\begin{figure}
\centering
\includegraphics[scale=0.6]{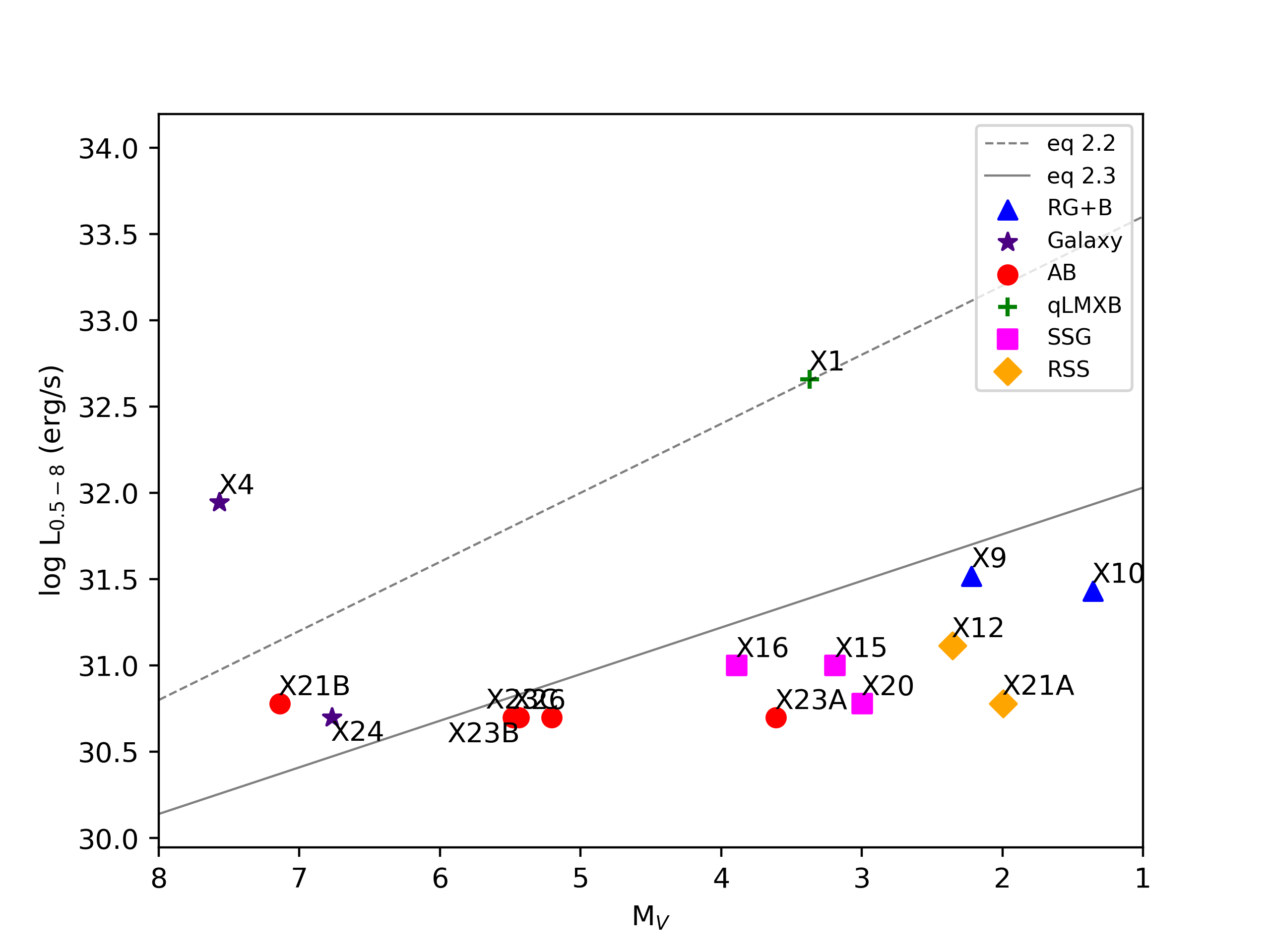}
\caption{X-ray luminosity as a function of absolute visual magnitude for the interesting stars given in Table \ref{tab:interesting_stars}. Equations 2.2 and 2.3 are taken from \citet{Verbunt08}.}
\label{fig:FxFv}
\end{figure}

X1 is the brightest X-ray source in NGC 362 ($L_{X(0.5-10)}$ = 4.2$\times10^{32}$ erg/s), and it is well-fit by the NS hydrogen atmosphere model (see section \ref{sec:3.3}).  This indicates that X1 is a quiescent low-mass X-ray binary (qLMXB). 
We identify a potentially interesting candidate star within the error circle, which falls within the SSG region in the $V_{606}-I_{814}$ CMD (Fig. \ref{fig:vi}). As the environment is rather crowded (see figure A1), it is possible that the photometry is affected. While SSG companion stars have not previously been identified in globular cluster qLMXBs, at least two redback MSPs in NGC 6397 show similar CMD positions \citep{Ferraro01,Grindlay01b,Zhao22,Zhang22}, and qLMXB accretion discs could be faint enough not to show up against a relatively bright star. Alternatively, the true qLMXB counterpart may be fainter (or obscured; Fig.~A1 shows that one other star within the error circle is not detected in HUGS) and lacking HUGS photometry.


X4's 
candidate counterpart lies near the main-sequence in the $V_{606}-I_{814}$ CMD, but is 1.5 magnitudes bluer than the main sequence in the $U_{336}-B_{438}$ and $UV_{275}-U{336}$ CMDs, suggesting it might be a CV. However, X4 has a zero membership probability in the \textit{HUGS} database, due to its distinctly different proper motion (visible in Figure A1, compared to the cluster stars). X4's proper motion in the \citet{Libralato22} {\it HUGS} proper motion database is the opposite of NGC 362's motion \citep{Gaia18glob}, verifying that X4 is a background active galactic nucleus (AGN). 


X9 and X10 have candidate counterparts that are located slightly to the right of the red giant branch in $V_{606}-I_{814}$, on or just to the left in $U_{336}-B_{438}$, and significantly blueward ($\sim$0.5 magnitudes) in $UV_{275}-U{336}$ (we label them "RG+B" in Table~\ref{tab:interesting_stars}). X-ray counterparts with such characteristics have not been identified in globular clusters previously, to our knowledge. We interpret the colour evolution to indicate a combination of a red straggler star (a giant with unusual evolution) along with a bright, blue source of light. The blue light could be produced by either a blue straggler star (suggesting the system is similar to Algol), or an accreting white dwarf star, in which case the system would be either a long-period CV, if the giant fills its Roche lobe (similar to AKO9 in 47 Tuc, \citealt{Knigge03}), or a symbiotic, if it accretes from a wind. The $L_X$/$M_V$ plot does not provide evidence against any of these possibilities (the X-rays could be produced by accretion onto a white dwarf, or by coronal emission from the giant).  

However, additional evidence comes from the variability study of \citet{Rozyczka16}, who associated X10 (in the Chandra Source Catalogue as X010322.09-705044.7) with their variable V24, an eclipsing binary of period 8.14 days. They obtained radial velocity measurements of V24, identifying it as a pair of 0.8 $\Msun$ stars with similar radial velocities. However, they also observed a highly variable lightcurve. The primary eclipse (its length implies the giant is filling its Roche lobe) is stable, but the secondary eclipse width varies by a factor of three on a timescale of years; \citet{Rozyczka16} suggest an accretion disk is sometimes present.  The deeper, stable eclipse must be the eclipse of the bluer object by the giant star. Optical spectroscopy could verify whether an accreting white dwarf is present (some high-accretion-rate CVs show absorption-line spectra, e.g. \citealt{Beuermann92}), or the accreting blue object is a blue straggler star. 


The candidate counterpart to X12, and the brighter candidate counterpart to X21 (labeled X21A), both appear to the right of the red giant branch in $V_{606}-I_{814}$, and are consistent with the red giant branch in $U_{336}-B_{438}$. X12 also appears redward of the red giant branch in $UV_{275}-U{336}$, while X21A falls on this branch in $UV_{275}-U{336}$. We identify these counterparts as red straggler stars, with X-rays likely powered by coronal activity; they are consistent with the AB region in the $L_X$/$M_V$ plot. 

X15, X16, and X20 are candidates for sub-subgiants, as previously discussed. All three sources have candidate counterparts slightly lower and to the right of the subgiant branch in all three CMDs, with X16 showing a colour bluer than the other two in $U_{336}-B_{438}$, but redder than them in  $UV_{275}-U{336}$ and $V_{606}-I_{814}$, suggestive of variability during the HUGS observations.

The candidate counterpart for X24 has a very blue $UV_{275}-U{336}$ colour, shifting redwards towards the main sequence in $V_{606}-I_{814}$, which could suggest a  CV nature. However, the \textit{HST} images show that this blue object is significantly extended (see Figure~A4), indicating that it is a background galaxy. 
The HUGS catalogue does not provide a proper motion or membership probability for this object, likely due to its extended nature. 

We identify five candidate ABs as potential counterparts to X21 (this we label X21B, with the X21A candidate a red straggler), X26, and X23 with three potential counterparts (X23A, X23B, X23C). All lie above the main sequence in $V_{606}-I_{814}$, with the candidate X23A lying $>1$ magnitude above the main sequence, and the rest lying $<$0.75 magnitudes above it. The candidates for X17 and X23B lie off the main sequence in $UV_{275}-U{336}$ also (X23B lying a magnitude off), while most lie on the main sequence in $UV_{275}-U{336}$ and all lie on the main sequence in $U_{336}-B_{438}$ (this CMD is significantly more vertical at the relevant magnitudes than the others). All but X21B lie below the lower empirical \citet{Verbunt08} equation setting an upper limit on AB $L_X$ values (Fig.~\ref{fig:FxFv}), and all lie below the higher empirical \citet{Verbunt08} upper limit, making them plausible AB candidates.






\section{Chance coincidence calculation}
\begin{figure*}
\centering
\includegraphics[scale=0.3]{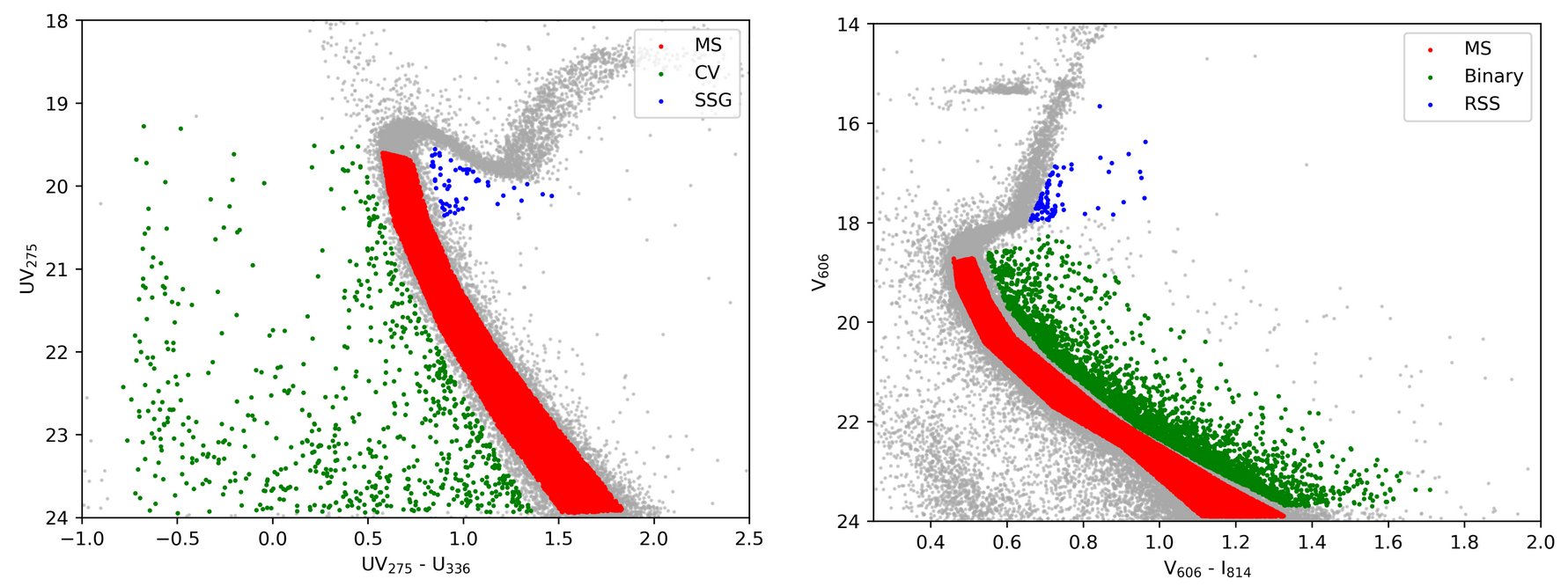}
\caption{CMDs showing the selected regions for main-sequence (MS), cataclysmic variable (CV), sub-subgiant (SSG), red straggler star (RSS), and binary sequence (Binary) stars in $UV_{275}-U_{336}$ and $V_{606}-I_{814}$ CMDs.}
\label{fig:chance_coincidence_cmds}
\end{figure*}

To identify how many of our identifications may be expected to be produced by chance, we select regions of the $UV_{275}-U_{336}$ and $V_{606}-I_{814}$ CMDs where we expect to find different types of X-ray counterparts (Fig.~\ref{fig:chance_coincidence_cmds}). We select regions in the $UV_{275}-U_{336}$ CMD to identify CVs (blueward of the main sequence) and sub-subgiants (below the subgiant branch). In the $V_{606}-I_{814}$ CMD we select regions for ABs (above the main sequence) and red stragglers (right of the giant branch). Sub-subgiant and AB identifications can overlap, as may some others. As a control, we also count the number of main-sequence (MS) stars in both filters. Finally, we identify stars that lie in the CV region in $UV_{275}-U{336}$ but on the MS in $V_{606}-I_{814}$ (which match our expectation that CVs will typically have blue and red light components). 

The density of stars varies substantially within our considered region, and X-ray sources are concentrated in the core. 
We therefore count the numbers of these stars within three spatial regions: the core; an annulus of 1.0 to 1.5 core radii; and an annulus from 1.5 core radii to 1 arcminute. Finally, we count the numbers of stars within the X-ray error circles (within these 3 regions) fitting these criteria.  
The results of these calculations are in Table~\ref{tab:chance_coincidence}. 

We note that we do not identify and discuss (in \ref{sec:4.1}) many of the stars which lie within X-ray error circles and also have interesting colours (e.g. the CV regions), as these stars typically appear crowded in one or more images, often with a neighbouring star which also has unusual colours (indicating that the {\it HUGS} photometry has shifted flux between the stars in different filters). Thus, our chance coincidence calculations will overstate the likelihood of a particular object being a spurious match to an X-ray source, as we cannot sift out all stars that appear crowded in all CMDs. Nevertheless, this exercise is useful to roughly estimate the degree of chance coincidences.

\begin{table*}
\centering
 \caption{Chance Coincidence results. Observed numbers are taken from the error circles, while predicted numbers are calculated from the relevant regions and area ratios, selected  using TOPCAT. Region 1 includes the core radius. Region 2 is an annulus from the core radius to 1.5 times the core radius. Region 3 is an annulus from 1.5 times the core radius to 1 arcminute.}
\begin{tabular}{lcccccc} 
 \hline
  Type &
      \multicolumn{3}{c}{Observed} &
      \multicolumn{3}{c}{Predicted} \\
     &  Region 1 & Region 2 & Region 3 & Region 1 & Region 2 & Region 3  \\ 
    \hline
CV in UV-U             & 4         & 2        & 2        & 3.9      & 1.4      & 0.7      \\
CV in UV-U \& MS in V-I    & 2         & 0        & 1        & 1.2      & 0.4      & 0.2      \\
Binary in V-I          & 28        & 15       & 5        & 22.7     & 10.3     & 4.2      \\
RSS in V-I             & 2         & 0        & 1        & 1.3     & 0.2      & 0.1      \\
MS in V-I              & 58       & 50       & 58       & 89.2    & 63.1     & 80.3     \\
SSG in U-V             & 2         & 1        & 1        & 0.8      & 0.1      & 0   \\
\hline
\end{tabular}
 \label{tab:chance_coincidence}
\end{table*}

First, the numbers of MS stars predicted and observed in the three regions are similar, though not identical (the predicted numbers average 1.4 times larger within each region). This suggests that the chance coincidence predictions give small overestimates, and should be taken as upper limits. 

We see that the numbers of objects observed in X-ray error circles in the CV region in $UV_{275}-U_{336}$ ("Observed") in Table~\ref{tab:chance_coincidence}) are similar to those predicted based on the CMD ("Predicted"), suggesting these could all be chance coincidences. However, when restricting consideration to objects in the CV region in $UV_{275}-U_{336}$ and on the MS in $V_{606}-I_{814}$, only 0.2 are predicted in region 3, where 1 is seen (X24). We are confident that both X24 and X4 (which lies just off the MS in $V_{606}-I_{814}$) are AGN, in the outer cluster region.

Stars in the AB region in $V_{606}-I_{814}$ are more likely to be observed than is predicted, in all three regions.
As the numbers of chance coincidences are predicted to be substantial (23 in the core), we cannot be highly confident that any of our matches with AB candidates are secure. However, the facts that we identify a number of good candidates, and that the predicted number of chance coincidences is less than we see, suggests that several of these ABs are real counterparts. 

Finally, the predicted numbers of spurious matches to red straggler stars and sub-subgiants are quite small, though the predictions are of order one false match of each within the core. Subtracting the predicted from observed values, we anticipate 1-2 real red straggler matches, and 3 real sub-subgiant matches in NGC 362, which happens to match the numbers of candidate counterparts that we advance in \ref{sec:4.1}. 
We conclude that most of our red straggler and sub-subgiant candidate counterparts are likely to be real counterparts.

\section{Discussion}

As is clear from Fig. 1, NGC 362 is rich in X-ray sources, which are concentrated into the cluster core. Although we have detected 33 sources within 1' of the cluster (1.2 half-mass radii; note that 4 of these lie outside the current best-estimate half-mass radius), excess X-ray emission within the cluster core can be seen in Figure \ref{fig:clusterxrayimg}. 
We estimate a total of 85 counts of excess emission (0.5-8 keV) in the cluster core (outside our 1" source regions, subtracting background and the wings of the point-spread-function for known sources), which (assuming the same spectrum as our faint sources) would convert to $L_X$(0.5-10 keV)$=8.5\times10^{31}$ erg/s. 
We estimate that our X-ray detection is complete down to about 10 counts, or $L_X$(0.5-10 keV)$=10^{31}$ erg/s, above which we find 16 sources. 

\subsection{Radial distributions}
We can use the radial distribution of X-ray sources to estimate the number of X-ray sources that are background objects, and (assuming that the sources have gravitationally relaxed) to infer the average mass of the cluster X-ray sources. 
Our approach has most recently been applied to M4, as described by \citet{Lugger23}. This analysis assumes that the radial profile of the cluster is reasonably well described by a \citet{King66} model. This has been demonstrated for NGC~362 by a number of studies, e.g.\ \citet{Trager95}, and we have verified it from the HUGS star counts for stars near the main-sequence turnoff (MSTO) mass. A key input to the X-ray source mass determination is the core radius value from the optical surface brightness profile fit. There is good agreement on this number in the literature. The published values are 10.2 arcsec \citep{Trager95}, 10.8 arcsec \citep[][2010 online version]{Harris96}, and 10.1 arcsec \citep{Baumgardt23}. Since the \citet{Harris96} value is based on \citet{Trager95}, we adopt a value of 10.2 arcsec. 

The analysis method used by \citet{Lugger23} computes the ratio of the typical mass $m_X$ of an X-ray source to that of a MSTO star $m_\textsc{msto}$ by a maximum likelihood fit of a generalised King model
\begin{equation}
    \label{eqn:profile}
    S(r) = S_0 \left[ 1 + \left(\frac{r}{r_c}\right)^2 \right]^{(1-3q_X)/2}
\end{equation}
to the surface density profile of the X-ray sources, where $q_X \equiv m_X/m_\textsc{msto}$ \citep[see][]{Heinke05}. In order to do this, it is necessary to correct for the extragalactic background X-ray source contribution. We use the extragalactic source counts of \citet{Giacconi01}, their eqn.~1, which gives an estimated background level of 0.69~arcmin$^{-2}$ for a 5-count source detection threshold. 

\begin{figure}
\centering
\includegraphics[width=\columnwidth]{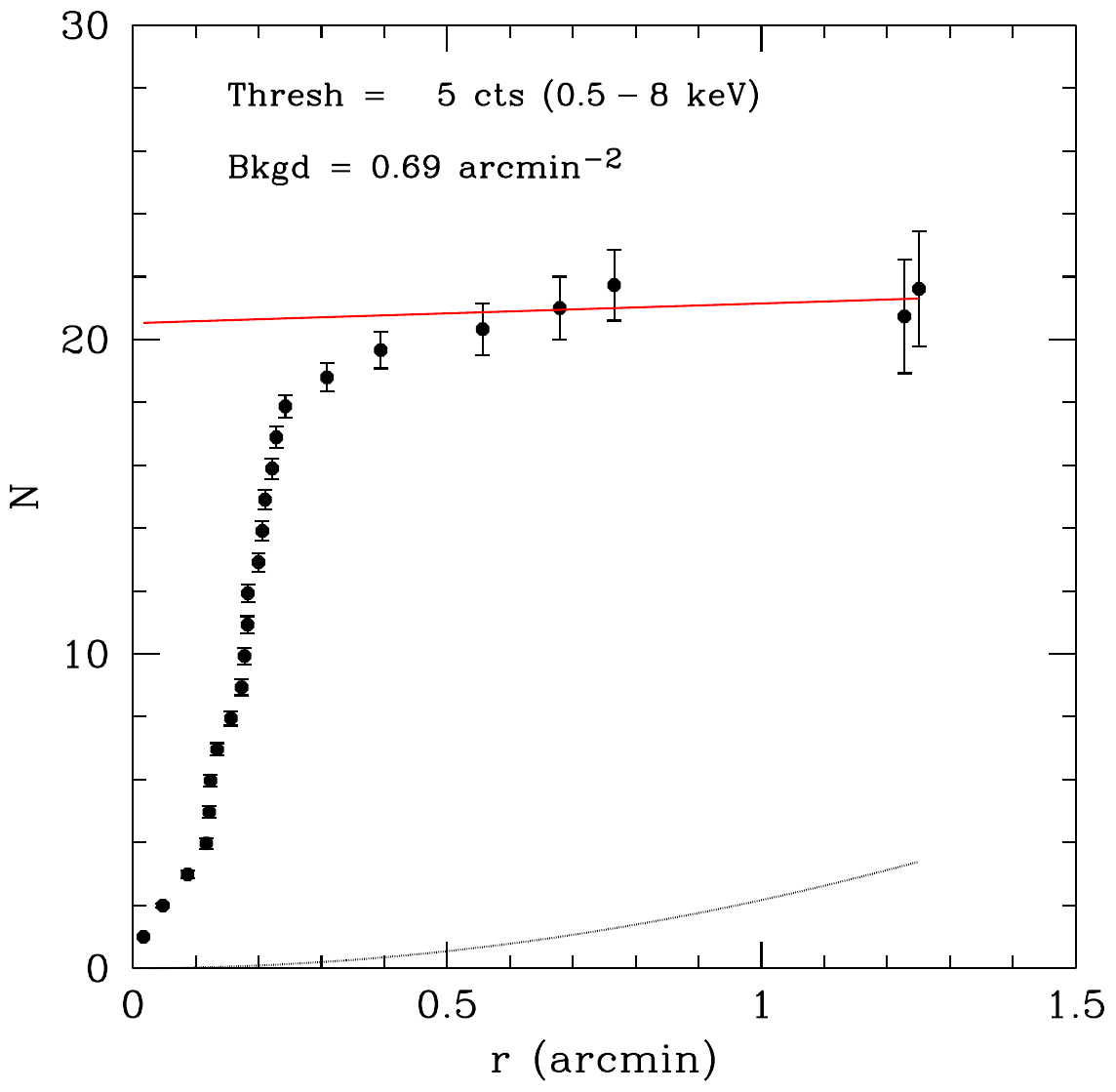}
\caption{Cumulative radial distribution of excess source counts over the background predicted from the \citet{Giacconi01} extragalactic source counts. Error bars represent the statistical uncertainty of the background correction. The dashed curve is the cumulative background distribution for the expected background density of 0.69 arcmin$^{-2}$ The red line is a least-squares fit to the profile for $r \ge 0.5 \,\mathrm{arcmin}$.}
\label{fig:cumulative_profile}
\end{figure}

Figure \ref{fig:cumulative_profile} shows the cumulative excess source number over the estimated background as function of the radial distance from the cluster centre; we use the centre determination by \citet{Goldsbury10}. As expected, the inferred cluster source distribution (excess over the background) becomes asymptotically flat beyond about 2 core radii. Thus, there is no evidence for an excess source population in the cluster halo, and this suggests that most of the eight X-ray sources farther than 2 core radii are background. Among these sources, we indeed identify X4 and X24 with background AGN, but we identify X10 with a cluster member at high confidence.

For the adopted parameter values of $r_c = 10.2^{\prime\prime}$, $n_\mathrm{bkgd} = 0.69\,\mathrm{arcmin}^{-2}$, and a 5-count detection threshold, the maximum-likelihood mass ratio is $q_X = 1.72 \pm 0.21$. This is in good agreement with the well-determined value of $q_X = 1.63 \pm 0.11$ that we obtained for 47~Tuc in \citet{Heinke05}. This mass ratio value also agrees with the values we have found for M71 \citep[$1.36 \pm 0.47$;][]{Elsner08}, M4 \citep[$1.53 \pm 0.24$;][]{Lugger23}, and M30 \citep[$1.61 \pm 0.20$;][]{Lugger07}.

For an assumed MSTO mass of $0.80\,\Msun$, the inferred typical \emph{Chandra} source mass in NGC~362 is $1.38 \pm 0.17\,\Msun$. This is a reasonable result for a mixture of ABs, CVs, MSPs, and qLMXBs, all of which should be somewhat more massive on average than a typical turnoff star \citep{Heinke05}.

\subsection{Comparison to other clusters}
NGC 362 happens to lie in a similar direction on the sky as 47 Tucanae (47 Tuc), though it is roughly twice as far away.  47 Tuc is 3.4 times as massive as NGC 362, but as NGC 362 has a denser core, NGC 362 has a stellar encounter rate 0.74$\pm0.20$ that of 47 Tuc \citep{Bahramian13}. It is interesting to compare their X-ray source content; above $L_X=10^{32}$ erg/s, 47 Tuc has 6 X-ray sources (3  qLMXBs, 2 CVs, and an AB), while NGC 362 has 3 (1 qLMXB; 2 unknown). Above $L_X$(0.5-6 keV)$=10^{31}$ erg/s, 47 Tuc has 39 total sources (from \citealt{Heinke05}, but assuming a 4.5 kpc distance, and separating MSPs 47 Tuc-F and 47 Tuc-S into two sources), while NGC 362 has 16 (one of those a background AGN). 
The total numbers of X-ray sources above $L_X=10^{31}$ erg/s are  lower than we might expect based purely on stellar encounter rates--38\% as many as in 47 Tuc, when we might expect 74$\pm20$\%--which may reflect other factors (metallicity, destruction of binaries in denser clusters, etc), or could be just small-number statistics. 

47 Tuc's sources above $10^{31}$ erg/s include 6 qLMXBs, 21 CVs (or candidates), 4 ABs, and 4 MSPs (or candidates). Thus, we may predict that if NGC 362 is similar, the majority of its sources above $10^{31}$ erg/s should be CVs. However, we have not identified any good candidate CVs among our X-ray sources. This at first seems odd, since we have the X-ray sensitivity (down to $L_X=10^{31}$ erg/s) and ultraviolet sensitivity (e.g. a $U_{336}-B_{438}$ CMD reaching over 4 magnitudes below the turnoff) that would be sufficient to identify a dozen CVs in 47 Tuc (see \citealt{Edmonds03a}), and thus we would predict identification of $\sim$8 CVs in this range in NGC 362. 
We do identify 2 AGN in the relevant CMD spaces, indicating that we have sensitivity to such sources in the cluster outskirts. However, the core is substantially more crowded especially in the optical. Fig.~\ref{fig:mr_rad} shows that the {\it HUGS} photometry is highly incomplete even a few magnitudes below the turnoff (visible at $M_R\sim3$). As inclusion in the {\it HUGS} photometry database relies on detection in $V_{606}$ or $I_{814}$, most CVs, which are likely to be located within $\sim$15" of the centre and be fainter than $M_R\sim6$, are likely to be missed. Future work may identify the likely CV population  
through a new photometric analysis of archival UV and $H-\alpha$ HST imaging  
of X-ray error circles. 

\begin{table*}
\centering
\begin{tabular}{lccccccc} 
 \hline
  Cluster & Distance & $r_h$  &
      $\Gamma$ & $\Gamma$ err & \multicolumn{2}{c}{X-ray srcs,  $L_X>10^{31}$ erg/s} & Refs \\
     & kpc & arcmin  & &  & Total & Fg/Bg (pred) &   \\ 
    \hline
Terzan 5 & 6.6 & 0.72 &  6800 & 1000 & 102 & 33 &  1,2  \\
NGC 6266 & 6.4 & 0.92 & 1670 & 710 & 39 & 2 & 3 \\
47 Tuc & 4.52 & 3.17 & 1000 & 15 & 33 & 1 & 4  \\
NGC 2808 & 10.1 & 0.80 & 923 & 67 & 16 & 1 & 5 \\
NGC 6388 & 11.2 & 0.52 & 900 & 240 & 52 & 1 & 6 \\
NGC 362 & 8.8 & 0.82 & 740 & 140 & 15 & 0 & 7 \\
NGC 6626 & 5.4 & 1.97 & 650 & 80 & 30 & 12 & 1,8  \\
NGC 6093 & 10.3 & 0.61 & 530 & 60 & 18 & 1 & 9 \\
NGC 6752 & 4.13 & 1.91 & 400 & 180 & 7 & 0 & 10 \\
NGC 7099 & 8.5 & 1.03 & 320 & 120 & 8 & 0 & 11, 12 \\
NGC 5904 & 7.5 & 1.77 & 160 & 40 & 12 & 3 & 1 \\
NGC 5139 & 5.43 & 5.00 & 90 & 27 & 32 & 14 & 13,14,15 \\
NGC 6397 & 2.48 & 2.90 & 84 & 18 & 9 & 1 & 16 \\
NGC 6656 & 3.30 & 3.36 & 78 & 32 & 8 & 3 & 1 \\
NGC 6121 & 1.85 & 4.33 & 27 & 12 & 1 & 0 & 17  \\
\hline
\end{tabular}
 \caption{Information about several globular clusters, including predicted stellar encounter rates \citep{Bahramian13} and observed numbers of X-ray sources, divided into likely members and nonmembers.  We use \citet{Baumgardt21} for distances, and \citet[][2010 revision]{Harris96} for half-mass radii and extinctions. 
 References: 1: \citet{Bahramian20}, 2: \citet{Cheng19}, 3: \citet{Oh20}, 4: \citet{Heinke05}, 5: \citet{Servillat08}, 6: \citet{Maxwell12}, 7: this work, 8: \citet{Cheng20}, 9: \citet{Heinke03c}, 10: \citet{Cohn21}, 11: \citet{Lugger07}, 12: \citet{Zhao20}, 13: \citet{Haggard09}, 14: \citet{Cool13}, 15: \citet{Henleywillis18}, 16: \citet{Bogdanov10}, 17: \citet{Bassa04}. }
 \label{tab:other_clusters}
\end{table*}

We also compare NGC 362 to other clusters with significant X-ray source populations. We collect information from 14 other clusters with published surveys down to $L_X$(0.5-10 keV)$=10^{31}$ erg/s, which we list in Table~\ref{tab:other_clusters}. If the published paper quoted a different energy range, we converted fluxes assuming a power-law of photon index 1.7. 
For most of the clusters, we estimate the number of background AGN based on the log N-log S estimate of \citet{Giacconi01}.  In a few clusters (NGC 6397, NGC 6752, NGC 6121), every source in this $L_X$ range has been identified.   However, in some clusters, especially at low Galactic latitude, the dominant contaminant is Galactic stars. For Terzan 5 and M28, we use estimates of the Galactic source contamination rate by \citet{Cheng19,Cheng20}. 

\begin{figure}
\centering
\includegraphics[width=1.06\columnwidth]{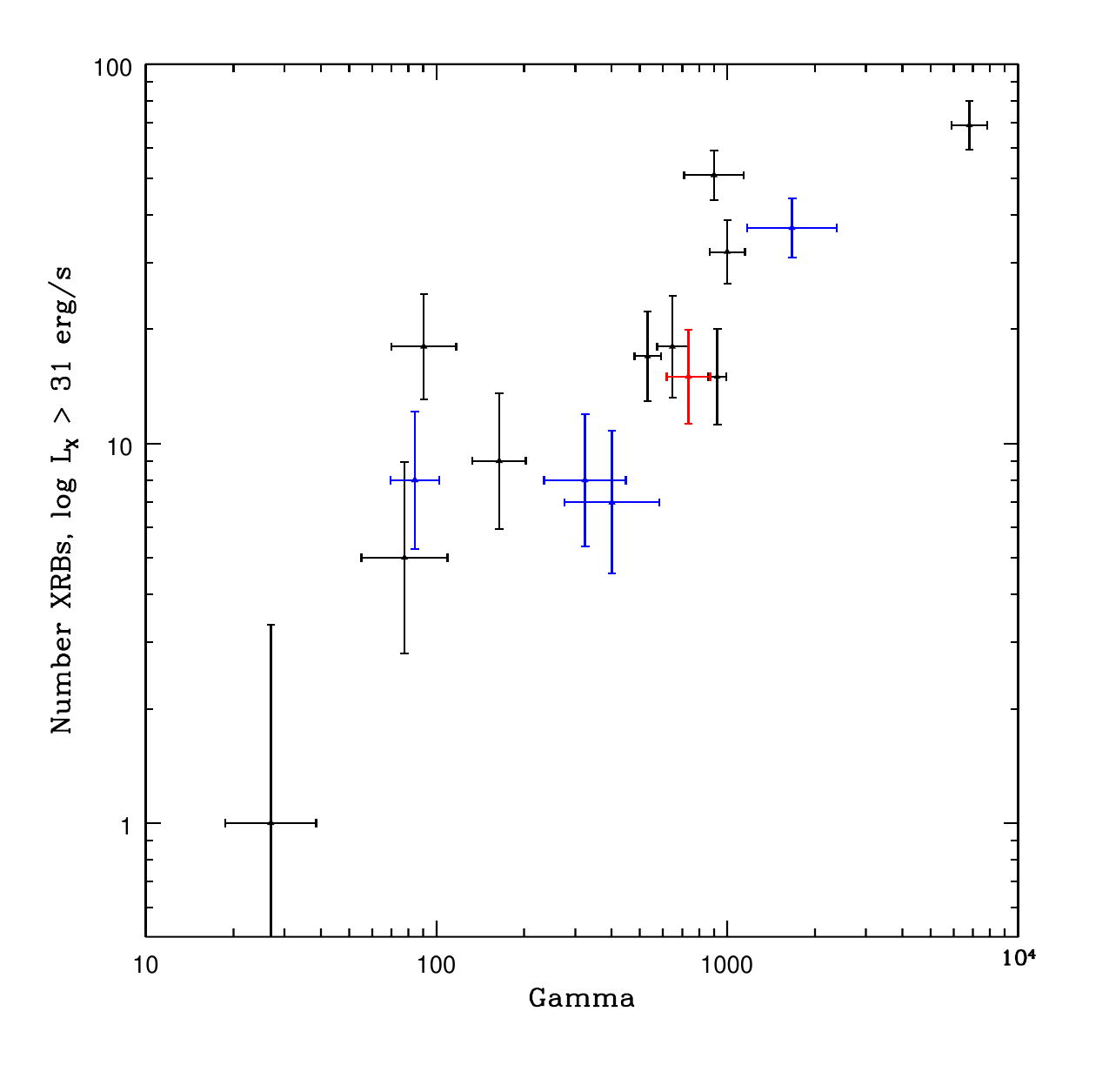}
\caption{Stellar encounter rates from \citet{Bahramian13} vs. observed numbers  of X-ray sources with $L_X$(0.5-10 keV)$>10^{31}$ erg/s (see Table~\ref{tab:other_clusters}). NGC 362 is in red, while other core-collapsed clusters are indicated in blue.}
\label{fig:compare}
\end{figure}

\begin{figure}
\centering
\includegraphics[width=\columnwidth]{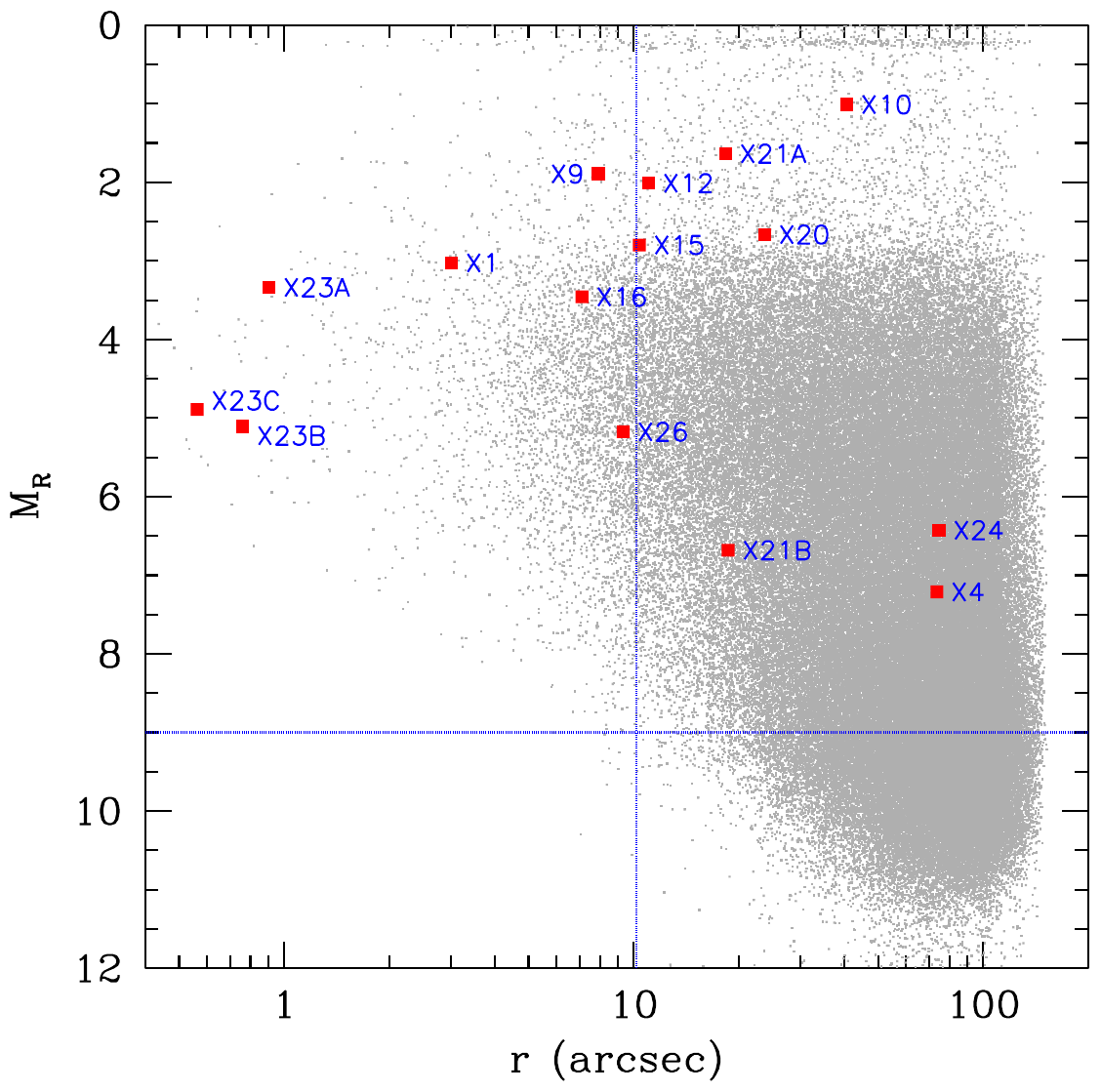}
\caption{Radial distribution, relative to NGC 362's center, of {\it HUGS} stars, vs.\ their estimated $M_R$ absolute magnitude (using $m_V-M_V=14.725$, \citealt[][2010 revision]{Harris96}, and assuming $R=(V+I)/2$). The vertical blue line represents the core radius. The horizontal blue line represents the absolute magnitude separating the bright and faint CV populations in NGC~6752 \citep{Cohn21}. The core of NGC 362 is severely incomplete at fainter magnitudes due to crowding.}
\label{fig:mr_rad}
\end{figure}

We plot the stellar encounter rates with errors from \citet{Bahramian13}, vs. the observed numbers of X-ray sources with $L_X$(0.5-10 keV)$>10^{31}$ erg/s, in Figure~\ref{fig:compare}. We calculate \citet{Gehrels86} errors on the total number of sources. NGC 362 is labeled in red, while other clusters marked as core-collapsed or possibly core-collapsed by \citet{Harris96} are indicated in blue. NGC 362's X-ray population appears similar to those of other clusters with similar encounter rates.

\subsection{Interesting objects}
The three sub-subgiants that we confidently identify have similar X-ray luminosities (0.5-10 keV $L_X$ of $6\times10^{30}$ to $10\times10^{30}$ erg/s) as the four sub-subgiant coronally active systems identified in 47 Tuc \citep{Albrow01,Heinke05}, which have $L_X$(0.5-6 keV) of $4.5\times10^{30}$ to $8.2\times10^{30}$ (excluding the more X-ray luminous CV AKO9). 
The qLMXB X1 may also be associated with a sub-subgiant optical counterpart, which would suggest an orbital period of 10-20 hours. 
Although most sub-subgiant systems are likely powered by coronal activity, placement in the sub-subgiant location in some CMDs may be a general feature of close-orbit nuclear evolution of close binaries, as illustrated by the variety of objects found there, including redback MSPs \citep{Ferraro01}, CVs \citep{Knigge03}, black hole candidates \citep{Shishkovsky18}, as well as both short-orbit and longer-orbit binaries composed of non-degenerate stars \citep{Mathieu03}.

 X9 and X10 show evidence of both red and blue components to their light. They could be Algol-type active binaries (where the stars include a red giant and a blue straggler), CVs containing a red giant filling its Roche lobe, or symbiotic stars.  
 No symbiotic stars have yet been verified in any globular cluster, though there is a candidate in $\omega$ Cen; \citep{Henleywillis18}. X10's optical lightcurve, presented by \citet{Rozyczka16}, shows eclipses with an 8.14 day period and large variability, suggesting a variable accretion disk.

\section{Conclusions}

Our analysis of the X-ray sources in NGC 362 finds 33 sources down to a limiting $L_X$(0.5-10 keV)$=3\times10^{30}$ erg/s, likely complete to $L_X=10^{31}$ erg/s. The numbers of X-ray sources are slightly lower than expected if scaling from 47 Tuc's X-ray content based on their relative stellar encounter rate, and their radial distribution is concentrated towards the core as appropriate for a stellar population with mass of 1.38$\pm0.17$ $\Msun$, consistent with known X-ray source populations. We identify the brightest X-ray source, X1, as a qLMXB based on its distinctive X-ray spectrum.  NGC 362's X-ray source characteristics are not surprising compared to similar globular clusters.

The optical/UV counterpart source population, however, does offer surprises. We do not identify any compelling CV candidates, 
although extrapolation by stellar encounter rate from 47 Tuc would suggest of order 8 CVs may exist among our detected X-ray sources. 
We infer that our lack of identified CVs is due to optical crowding in the core.  This suggests that CVs may be identified using a new reduction of the deep HST F225W UV imaging of this cluster that is available in the archive, since the HUGS source list was prepared using less deep, longer wavelength data.
We find several plausible AB candidate counterparts. 
On the other hand, we find numerous relatively bright likely optical counterparts with unusual CMD positions; 3 plausibly coronally-powered sub-subgiants, a possible sub-subgiant optical counterpart to the qLMXB, 2 objects in the red straggler location in all CMDs, and 2 objects that lie in the red straggler location in our optical CMD but shift well to the blue of the giant branch in ultraviolet CMDs. Chance coincidence tests indicate that few if any of these are false matches, and one of the last group (X10) has an optical lightcurve showing eclipses on an 8.14 day orbit with strong variability indicative of a variable accretion disk around the bluer star \citep{Rozyczka16}. The majority of these bright optical counterparts are likely to be powered by coronal activity, but other explanations are possible. Optical spectroscopy of these various bright optical X-ray counterparts could be very helpful in understanding their nature.

\section*{Acknowledgements}
GK is supported by a Mitacs Globalink Research Internship. 
CH is supported by NSERC Discovery Grant RGPIN-2023-04264.

\section*{Data Availability}

The {\it Chandra} data used in this work is available via the {\it Chandra} Data Archive, at \url{http://cda.harvard.edu/chaser/}, while the {\it HST} HUGS catalogues are available at \url{https://archive.stsci.edu/prepds/hugs/}.


\bibliographystyle{mnras}
\bibliography{references} 

\begin{thebibliography}{}
\makeatletter
\relax
\def\mn@urlcharsother{\let\do\@makeother \do\$\do\&\do\#\do\^\do\_\do\%\do\~}
\def\mn@doi{\begingroup\mn@urlcharsother \@ifnextchar [ {\mn@doi@} {\mn@doi@[]}}
\def\mn@doi@[#1]#2{\def\@tempa{#1}\ifx\@tempa\@empty \href {http://dx.doi.org/#2} {doi:#2}\else \href {http://dx.doi.org/#2} {#1}\fi \endgroup}
\def\mn@eprint#1#2{\mn@eprint@#1:#2::\@nil}
\def\mn@eprint@arXiv#1{\href {http://arxiv.org/abs/#1} {{\tt arXiv:#1}}}
\def\mn@eprint@dblp#1{\href {http://dblp.uni-trier.de/rec/bibtex/#1.xml} {dblp:#1}}
\def\mn@eprint@#1:#2:#3:#4\@nil{\def\@tempa {#1}\def\@tempb {#2}\def\@tempc {#3}\ifx \@tempc \@empty \let \@tempc \@tempb \let \@tempb \@tempa \fi \ifx \@tempb \@empty \def\@tempb {arXiv}\fi \@ifundefined {mn@eprint@\@tempb}{\@tempb:\@tempc}{\expandafter \expandafter \csname mn@eprint@\@tempb\endcsname \expandafter{\@tempc}}}

\bibitem[\protect\citeauthoryear{{Albrow}, {Gilliland}, {Brown}, {Edmonds}, {Guhathakurta}  \& {Sarajedini}}{{Albrow} et~al.}{2001}]{Albrow01}
{Albrow} M.~D.,  {Gilliland} R.~L.,  {Brown} T.~M.,  {Edmonds} P.~D.,  {Guhathakurta} P.,   {Sarajedini} A.,  2001, \mn@doi [\apj] {10.1086/322353}, \href {https://ui.adsabs.harvard.edu/abs/2001ApJ...559.1060A} {559, 1060}

\bibitem[\protect\citeauthoryear{{Alpar}, {Cheng}, {Ruderman}  \& {Shaham}}{{Alpar} et~al.}{1982}]{Alpar82}
{Alpar} M.~A.,  {Cheng} A.~F.,  {Ruderman} M.~A.,   {Shaham} J.,  1982, \mn@doi [\nat] {10.1038/300728a0}, \href {https://ui.adsabs.harvard.edu/abs/1982Natur.300..728A} {300, 728}

\bibitem[\protect\citeauthoryear{{Anderson} et~al.,}{{Anderson} et~al.}{2008}]{2008AJ....135.2055A}
{Anderson} J.,  et~al., 2008, \mn@doi [\aj] {10.1088/0004-6256/135/6/2055}, \href {https://ui.adsabs.harvard.edu/abs/2008AJ....135.2055A} {135, 2055}

\bibitem[\protect\citeauthoryear{{Bagchi}, {Lorimer}  \& {Chennamangalam}}{{Bagchi} et~al.}{2011}]{Bagchi11}
{Bagchi} M.,  {Lorimer} D.~R.,   {Chennamangalam} J.,  2011, \mn@doi [\mnras] {10.1111/j.1365-2966.2011.19498.x}, \href {https://ui.adsabs.harvard.edu/abs/2011MNRAS.418..477B} {418, 477}

\bibitem[\protect\citeauthoryear{{Bahramian}, {Heinke}, {Sivakoff}  \& {Gladstone}}{{Bahramian} et~al.}{2013}]{Bahramian13}
{Bahramian} A.,  {Heinke} C.~O.,  {Sivakoff} G.~R.,   {Gladstone} J.~C.,  2013, \mn@doi [\apj] {10.1088/0004-637X/766/2/136}, \href {https://ui.adsabs.harvard.edu/abs/2013ApJ...766..136B} {766, 136}

\bibitem[\protect\citeauthoryear{{Bahramian}, {Heinke}, {Degenaar}, {Chomiuk}, {Wijnands}, {Strader}, {Ho}  \& {Pooley}}{{Bahramian} et~al.}{2015}]{2015MNRAS.452.3475B}
{Bahramian} A.,  {Heinke} C.~O.,  {Degenaar} N.,  {Chomiuk} L.,  {Wijnands} R.,  {Strader} J.,  {Ho} W. C.~G.,   {Pooley} D.,  2015, \mn@doi [\mnras] {10.1093/mnras/stv1585}, \href {https://ui.adsabs.harvard.edu/abs/2015MNRAS.452.3475B} {452, 3475}

\bibitem[\protect\citeauthoryear{{Bahramian} et~al.,}{{Bahramian} et~al.}{2020}]{Bahramian20}
{Bahramian} A.,  et~al., 2020, \mn@doi [\apj] {10.3847/1538-4357/aba51d}, \href {https://ui.adsabs.harvard.edu/abs/2020ApJ...901...57B} {901, 57}

\bibitem[\protect\citeauthoryear{{Bassa} et~al.,}{{Bassa} et~al.}{2004}]{Bassa04}
{Bassa} C.,  et~al., 2004, \mn@doi [\apj] {10.1086/421259}, \href {https://ui.adsabs.harvard.edu/abs/2004ApJ...609..755B} {609, 755}

\bibitem[\protect\citeauthoryear{{Bassa}, {Pooley}, {Verbunt}, {Homer}, {Anderson}  \& {Lewin}}{{Bassa} et~al.}{2008}]{Bassa08}
{Bassa} C.~G.,  {Pooley} D.,  {Verbunt} F.,  {Homer} L.,  {Anderson} S.~F.,   {Lewin} W.~H.~G.,  2008, \mn@doi [\aap] {10.1051/0004-6361:200809350}, \href {https://ui.adsabs.harvard.edu/abs/2008A&A...488..921B} {488, 921}

\bibitem[\protect\citeauthoryear{{Baumgardt} \& {Vasiliev}}{{Baumgardt} \& {Vasiliev}}{2021}]{Baumgardt21}
{Baumgardt} H.,  {Vasiliev} E.,  2021, \mn@doi [\mnras] {10.1093/mnras/stab1474}, \href {https://ui.adsabs.harvard.edu/abs/2021MNRAS.505.5957B} {505, 5957}

\bibitem[\protect\citeauthoryear{{Baumgardt}, {Sollima}, {Hilker}, {Bellini}, {Vasiliev}, {Henault-Brunet}  \& {Dickson}}{{Baumgardt} et~al.}{2023}]{Baumgardt23}
{Baumgardt} H.,  {Sollima} A.,  {Hilker} M.,  {Bellini} A.,  {Vasiliev} E.,  {Henault-Brunet} V.,   {Dickson} N.,  2023, Fundamental parameters of Galactic globular clusters (as of May 2021), \url{https://people.smp.uq.edu.au/HolgerBaumgardt/globular/}

\bibitem[\protect\citeauthoryear{{Bellini} et~al.,}{{Bellini} et~al.}{2009}]{2009A&A...493..959B}
{Bellini} A.,  et~al., 2009, \mn@doi [\aap] {10.1051/0004-6361:200810880}, \href {https://ui.adsabs.harvard.edu/abs/2009A&A...493..959B} {493, 959}

\bibitem[\protect\citeauthoryear{{Bellini}, {Anderson}, {Bedin}, {King}, {van der Marel}, {Piotto}  \& {Cool}}{{Bellini} et~al.}{2017}]{2017ApJ...842....6B}
{Bellini} A.,  {Anderson} J.,  {Bedin} L.~R.,  {King} I.~R.,  {van der Marel} R.~P.,  {Piotto} G.,   {Cool} A.,  2017, \mn@doi [\apj] {10.3847/1538-4357/aa7059}, \href {https://ui.adsabs.harvard.edu/abs/2017ApJ...842....6B} {842, 6}

\bibitem[\protect\citeauthoryear{{Belloni}, {Giersz}, {Rivera Sandoval}, {Askar}  \& {Cieciel{\r{a}}g}}{{Belloni} et~al.}{2019}]{Belloni19}
{Belloni} D.,  {Giersz} M.,  {Rivera Sandoval} L.~E.,  {Askar} A.,   {Cieciel{\r{a}}g} P.,  2019, \mn@doi [\mnras] {10.1093/mnras/sty3097}, \href {https://ui.adsabs.harvard.edu/abs/2019MNRAS.483..315B} {483, 315}

\bibitem[\protect\citeauthoryear{{Belloni}, {Miko{\l}ajewska}, {I{\l}kiewicz}, {Schreiber}, {Giersz}, {Rivera Sandoval}  \& {Rodrigues}}{{Belloni} et~al.}{2020}]{Belloni20}
{Belloni} D.,  {Miko{\l}ajewska} J.,  {I{\l}kiewicz} K.,  {Schreiber} M.~R.,  {Giersz} M.,  {Rivera Sandoval} L.~E.,   {Rodrigues} C.~V.,  2020, \mn@doi [\mnras] {10.1093/mnras/staa1714}, \href {https://ui.adsabs.harvard.edu/abs/2020MNRAS.496.3436B} {496, 3436}

\bibitem[\protect\citeauthoryear{{Beuermann}, {Stasiewski}  \& {Schwope}}{{Beuermann} et~al.}{1992}]{Beuermann92}
{Beuermann} K.,  {Stasiewski} U.,   {Schwope} A.~D.,  1992, \aap, \href {https://ui.adsabs.harvard.edu/abs/1992A&A...256..433B} {256, 433}

\bibitem[\protect\citeauthoryear{{Bogdanov}, {Grindlay}, {Heinke}, {Camilo}, {Freire}  \& {Becker}}{{Bogdanov} et~al.}{2006}]{Bogdanov06}
{Bogdanov} S.,  {Grindlay} J.~E.,  {Heinke} C.~O.,  {Camilo} F.,  {Freire} P. C.~C.,   {Becker} W.,  2006, \mn@doi [\apj] {10.1086/505133}, \href {https://ui.adsabs.harvard.edu/abs/2006ApJ...646.1104B} {646, 1104}

\bibitem[\protect\citeauthoryear{{Bogdanov}, {van den Berg}, {Heinke}, {Cohn}, {Lugger}  \& {Grindlay}}{{Bogdanov} et~al.}{2010}]{Bogdanov10}
{Bogdanov} S.,  {van den Berg} M.,  {Heinke} C.~O.,  {Cohn} H.~N.,  {Lugger} P.~M.,   {Grindlay} J.~E.,  2010, \mn@doi [\apj] {10.1088/0004-637X/709/1/241}, \href {https://ui.adsabs.harvard.edu/abs/2010ApJ...709..241B} {709, 241}

\bibitem[\protect\citeauthoryear{{Cadelano} et~al.,}{{Cadelano} et~al.}{2015}]{Cadelano15}
{Cadelano} M.,  et~al., 2015, \mn@doi [\apj] {10.1088/0004-637X/807/1/91}, \href {https://ui.adsabs.harvard.edu/abs/2015ApJ...807...91C} {807, 91}

\bibitem[\protect\citeauthoryear{{Chen} et~al.,}{{Chen} et~al.}{2023}]{Chen23}
{Chen} J.,  et~al., 2023, \mn@doi [\apj] {10.3847/1538-4357/acc583}, \href {https://ui.adsabs.harvard.edu/abs/2023ApJ...948...84C} {948, 84}

\bibitem[\protect\citeauthoryear{{Cheng}, {Li}, {Xu}  \& {Li}}{{Cheng} et~al.}{2018}]{Cheng18}
{Cheng} Z.,  {Li} Z.,  {Xu} X.,   {Li} X.,  2018, \mn@doi [\apj] {10.3847/1538-4357/aaba16}, \href {https://ui.adsabs.harvard.edu/abs/2018ApJ...858...33C} {858, 33}

\bibitem[\protect\citeauthoryear{{Cheng}, {Li}, {Fang}, {Li}  \& {Xu}}{{Cheng} et~al.}{2019}]{Cheng19}
{Cheng} Z.,  {Li} Z.,  {Fang} T.,  {Li} X.,   {Xu} X.,  2019, \mn@doi [\apj] {10.3847/1538-4357/ab3c6d}, \href {https://ui.adsabs.harvard.edu/abs/2019ApJ...883...90C} {883, 90}

\bibitem[\protect\citeauthoryear{{Cheng}, {Mu}, {Li}, {Xu}, {Wang}  \& {Li}}{{Cheng} et~al.}{2020}]{Cheng20}
{Cheng} Z.,  {Mu} H.,  {Li} Z.,  {Xu} X.,  {Wang} W.,   {Li} X.,  2020, \mn@doi [\apj] {10.3847/1538-4357/ab7933}, \href {https://ui.adsabs.harvard.edu/abs/2020ApJ...892...16C} {892, 16}

\bibitem[\protect\citeauthoryear{{Clark}}{{Clark}}{1975}]{Clark75}
{Clark} G.~W.,  1975, \mn@doi [\apjl] {10.1086/181869}, \href {https://ui.adsabs.harvard.edu/abs/1975ApJ...199L.143C} {199, L143}

\bibitem[\protect\citeauthoryear{{Cohn} et~al.,}{{Cohn} et~al.}{2021}]{Cohn21}
{Cohn} H.~N.,  et~al., 2021, \mn@doi [\mnras] {10.1093/mnras/stab2636}, \href {https://ui.adsabs.harvard.edu/abs/2021MNRAS.508.2823C} {508, 2823}

\bibitem[\protect\citeauthoryear{{Cool}, {Haggard}, {Arias}, {Brochmann}, {Dorfman}, {Gafford}, {White}  \& {Anderson}}{{Cool} et~al.}{2013}]{Cool13}
{Cool} A.~M.,  {Haggard} D.,  {Arias} T.,  {Brochmann} M.,  {Dorfman} J.,  {Gafford} A.,  {White} V.,   {Anderson} J.,  2013, \mn@doi [\apj] {10.1088/0004-637X/763/2/126}, \href {https://ui.adsabs.harvard.edu/abs/2013ApJ...763..126C} {763, 126}

\bibitem[\protect\citeauthoryear{{Edmonds}, {Gilliland}, {Heinke}, {Grindlay}  \& {Camilo}}{{Edmonds} et~al.}{2001}]{Edmonds01}
{Edmonds} P.~D.,  {Gilliland} R.~L.,  {Heinke} C.~O.,  {Grindlay} J.~E.,   {Camilo} F.,  2001, \mn@doi [\apjl] {10.1086/323122}, \href {https://ui.adsabs.harvard.edu/abs/2001ApJ...557L..57E} {557, L57}

\bibitem[\protect\citeauthoryear{{Edmonds}, {Gilliland}, {Heinke}  \& {Grindlay}}{{Edmonds} et~al.}{2003}]{Edmonds03a}
{Edmonds} P.~D.,  {Gilliland} R.~L.,  {Heinke} C.~O.,   {Grindlay} J.~E.,  2003, \mn@doi [\apj] {10.1086/378193}, \href {https://ui.adsabs.harvard.edu/abs/2003ApJ...596.1177E} {596, 1177}

\bibitem[\protect\citeauthoryear{{Elsner} et~al.,}{{Elsner} et~al.}{2008}]{Elsner08}
{Elsner} R.~F.,  et~al., 2008, \mn@doi [\apj] {10.1086/591899}, \href {https://ui.adsabs.harvard.edu/abs/2008ApJ...687.1019E} {687, 1019}

\bibitem[\protect\citeauthoryear{{Ferraro}, {Possenti}, {D'Amico}  \& {Sabbi}}{{Ferraro} et~al.}{2001}]{Ferraro01}
{Ferraro} F.~R.,  {Possenti} A.,  {D'Amico} N.,   {Sabbi} E.,  2001, \mn@doi [\apjl] {10.1086/324563}, \href {https://ui.adsabs.harvard.edu/abs/2001ApJ...561L..93F} {561, L93}

\bibitem[\protect\citeauthoryear{{Gaia Collaboration} et~al.,}{{Gaia Collaboration} et~al.}{2016}]{2016A&A...595A...2G}
{Gaia Collaboration} et~al., 2016, \mn@doi [\aap] {10.1051/0004-6361/201629512}, \href {https://ui.adsabs.harvard.edu/abs/2016A&A...595A...2G} {595, A2}

\bibitem[\protect\citeauthoryear{{Gaia Collaboration} et~al.,}{{Gaia Collaboration} et~al.}{2018}]{Gaia18glob}
{Gaia Collaboration} et~al., 2018, \mn@doi [\aap] {10.1051/0004-6361/201832698}, \href {https://ui.adsabs.harvard.edu/abs/2018A&A...616A..12G} {616, A12}

\bibitem[\protect\citeauthoryear{{Gaia Collaboration} et~al.,}{{Gaia Collaboration} et~al.}{2023}]{2023A&A...674A...1G}
{Gaia Collaboration} et~al., 2023, \mn@doi [\aap] {10.1051/0004-6361/202243940}, \href {https://ui.adsabs.harvard.edu/abs/2023A&A...674A...1G} {674, A1}

\bibitem[\protect\citeauthoryear{{Gehrels}}{{Gehrels}}{1986}]{Gehrels86}
{Gehrels} N.,  1986, \mn@doi [\apj] {10.1086/164079}, \href {https://ui.adsabs.harvard.edu/abs/1986ApJ...303..336G} {303, 336}

\bibitem[\protect\citeauthoryear{{Geller} et~al.,}{{Geller} et~al.}{2017}]{Geller17}
{Geller} A.~M.,  et~al., 2017, \mn@doi [\apj] {10.3847/1538-4357/aa6af3}, \href {https://ui.adsabs.harvard.edu/abs/2017ApJ...840...66G} {840, 66}

\bibitem[\protect\citeauthoryear{{Giacconi} et~al.,}{{Giacconi} et~al.}{2001}]{Giacconi01}
{Giacconi} R.,  et~al., 2001, \mn@doi [\apj] {10.1086/320222}, \href {https://ui.adsabs.harvard.edu/abs/2001ApJ...551..624G} {551, 624}

\bibitem[\protect\citeauthoryear{{Goldsbury}, {Richer}, {Anderson}, {Dotter}, {Sarajedini}  \& {Woodley}}{{Goldsbury} et~al.}{2010}]{Goldsbury10}
{Goldsbury} R.,  {Richer} H.~B.,  {Anderson} J.,  {Dotter} A.,  {Sarajedini} A.,   {Woodley} K.,  2010, \mn@doi [\aj] {10.1088/0004-6256/140/6/1830}, \href {http://adsabs.harvard.edu/abs/2010AJ....140.1830G} {140, 1830}

\bibitem[\protect\citeauthoryear{{Grindlay}, {Heinke}, {Edmonds}, {Murray}  \& {Cool}}{{Grindlay} et~al.}{2001}]{Grindlay01b}
{Grindlay} J.~E.,  {Heinke} C.~O.,  {Edmonds} P.~D.,  {Murray} S.~S.,   {Cool} A.~M.,  2001, \mn@doi [\apjl] {10.1086/338499}, \href {https://ui.adsabs.harvard.edu/abs/2001ApJ...563L..53G} {563, L53}

\bibitem[\protect\citeauthoryear{{G{\"u}del}}{{G{\"u}del}}{2004}]{Gudel04}
{G{\"u}del} M.,  2004, \mn@doi [\aapr] {10.1007/s00159-004-0023-2}, \href {https://ui.adsabs.harvard.edu/abs/2004A&ARv..12...71G} {12, 71}

\bibitem[\protect\citeauthoryear{{Guillot}, {Rutledge}, {Bildsten}, {Brown}, {Pavlov}  \& {Zavlin}}{{Guillot} et~al.}{2009}]{Guillot09}
{Guillot} S.,  {Rutledge} R.~E.,  {Bildsten} L.,  {Brown} E.~F.,  {Pavlov} G.~G.,   {Zavlin} V.~E.,  2009, \mn@doi [\mnras] {10.1111/j.1365-2966.2008.14076.x}, \href {https://ui.adsabs.harvard.edu/abs/2009MNRAS.392..665G} {392, 665}

\bibitem[\protect\citeauthoryear{{Haggard}, {Cool}, {Anderson}, {Edmonds}, {Callanan}, {Heinke}, {Grindlay}  \& {Bailyn}}{{Haggard} et~al.}{2004}]{Haggard04}
{Haggard} D.,  {Cool} A.~M.,  {Anderson} J.,  {Edmonds} P.~D.,  {Callanan} P.~J.,  {Heinke} C.~O.,  {Grindlay} J.~E.,   {Bailyn} C.~D.,  2004, \mn@doi [\apj] {10.1086/421549}, \href {https://ui.adsabs.harvard.edu/abs/2004ApJ...613..512H} {613, 512}

\bibitem[\protect\citeauthoryear{{Haggard}, {Cool}  \& {Davies}}{{Haggard} et~al.}{2009}]{Haggard09}
{Haggard} D.,  {Cool} A.~M.,   {Davies} M.~B.,  2009, \mn@doi [\apj] {10.1088/0004-637X/697/1/224}, \href {https://ui.adsabs.harvard.edu/abs/2009ApJ...697..224H} {697, 224}

\bibitem[\protect\citeauthoryear{{Harris}}{{Harris}}{1996}]{Harris96}
{Harris} W.~E.,  1996, \mn@doi [\aj] {10.1086/118116}, \href {https://ui.adsabs.harvard.edu/abs/1996AJ....112.1487H} {112, 1487}

\bibitem[\protect\citeauthoryear{{Harris}}{{Harris}}{2010}]{2010arXiv1012.3224H}
{Harris} W.~E.,  2010, \mn@doi [arXiv e-prints] {10.48550/arXiv.1012.3224}, \href {https://ui.adsabs.harvard.edu/abs/2010arXiv1012.3224H} {p. arXiv:1012.3224}

\bibitem[\protect\citeauthoryear{{Heinke}, {Grindlay}, {Lugger}, {Cohn}, {Edmonds}, {Lloyd}  \& {Cool}}{{Heinke} et~al.}{2003a}]{Heinke03d}
{Heinke} C.~O.,  {Grindlay} J.~E.,  {Lugger} P.~M.,  {Cohn} H.~N.,  {Edmonds} P.~D.,  {Lloyd} D.~A.,   {Cool} A.~M.,  2003a, \mn@doi [\apj] {10.1086/378885}, \href {https://ui.adsabs.harvard.edu/abs/2003ApJ...598..501H} {598, 501}

\bibitem[\protect\citeauthoryear{{Heinke}, {Grindlay}, {Edmonds}, {Lloyd}, {Murray}, {Cohn}  \& {Lugger}}{{Heinke} et~al.}{2003b}]{Heinke03c}
{Heinke} C.~O.,  {Grindlay} J.~E.,  {Edmonds} P.~D.,  {Lloyd} D.~A.,  {Murray} S.~S.,  {Cohn} H.~N.,   {Lugger} P.~M.,  2003b, \mn@doi [\apj] {10.1086/378884}, \href {https://ui.adsabs.harvard.edu/abs/2003ApJ...598..516H} {598, 516}

\bibitem[\protect\citeauthoryear{{Heinke}, {Grindlay}, {Edmonds}, {Cohn}, {Lugger}, {Camilo}, {Bogdanov}  \& {Freire}}{{Heinke} et~al.}{2005}]{Heinke05}
{Heinke} C.~O.,  {Grindlay} J.~E.,  {Edmonds} P.~D.,  {Cohn} H.~N.,  {Lugger} P.~M.,  {Camilo} F.,  {Bogdanov} S.,   {Freire} P.~C.,  2005, \mn@doi [\apj] {10.1086/429899}, \href {https://ui.adsabs.harvard.edu/abs/2005ApJ...625..796H} {625, 796}

\bibitem[\protect\citeauthoryear{{Heinke}, {Rybicki}, {Narayan}  \& {Grindlay}}{{Heinke} et~al.}{2006}]{2006ApJ...644.1090H}
{Heinke} C.~O.,  {Rybicki} G.~B.,  {Narayan} R.,   {Grindlay} J.~E.,  2006, \mn@doi [\apj] {10.1086/503701}, \href {https://ui.adsabs.harvard.edu/abs/2006ApJ...644.1090H} {644, 1090}

\bibitem[\protect\citeauthoryear{{Heinke} et~al.,}{{Heinke} et~al.}{2020}]{Heinke20}
{Heinke} C.~O.,  et~al., 2020, \mn@doi [\mnras] {10.1093/mnras/staa194}, \href {https://ui.adsabs.harvard.edu/abs/2020MNRAS.492.5684H} {492, 5684}

\bibitem[\protect\citeauthoryear{{Henleywillis}, {Cool}, {Haggard}, {Heinke}, {Callanan}  \& {Zhao}}{{Henleywillis} et~al.}{2018}]{Henleywillis18}
{Henleywillis} S.,  {Cool} A.~M.,  {Haggard} D.,  {Heinke} C.,  {Callanan} P.,   {Zhao} Y.,  2018, \mn@doi [\mnras] {10.1093/mnras/sty675}, \href {https://ui.adsabs.harvard.edu/abs/2018MNRAS.479.2834H} {479, 2834}

\bibitem[\protect\citeauthoryear{{Hertz} \& {Grindlay}}{{Hertz} \& {Grindlay}}{1983}]{1983ApJ...267L..83H}
{Hertz} P.,  {Grindlay} J.~E.,  1983, \mn@doi [\apjl] {10.1086/184008}, \href {https://ui.adsabs.harvard.edu/abs/1983ApJ...267L..83H} {267, L83}

\bibitem[\protect\citeauthoryear{{Humphrey}, {Liu}  \& {Buote}}{{Humphrey} et~al.}{2009}]{2009ApJ...693..822H}
{Humphrey} P.~J.,  {Liu} W.,   {Buote} D.~A.,  2009, \mn@doi [\apj] {10.1088/0004-637X/693/1/822}, \href {https://ui.adsabs.harvard.edu/abs/2009ApJ...693..822H} {693, 822}

\bibitem[\protect\citeauthoryear{{Kim} et~al.,}{{Kim} et~al.}{2007}]{2007ApJS..169..401K}
{Kim} M.,  et~al., 2007, \mn@doi [\apjs] {10.1086/511634}, \href {https://ui.adsabs.harvard.edu/abs/2007ApJS..169..401K} {169, 401}

\bibitem[\protect\citeauthoryear{{King}}{{King}}{1966}]{King66}
{King} I.~R.,  1966, \mn@doi [\aj] {10.1086/109857}, \href {https://ui.adsabs.harvard.edu/abs/1966AJ.....71...64K} {71, 64}

\bibitem[\protect\citeauthoryear{{Knigge}, {Zurek}, {Shara}, {Long}  \& {Gilliland}}{{Knigge} et~al.}{2003}]{Knigge03}
{Knigge} C.,  {Zurek} D.~R.,  {Shara} M.~M.,  {Long} K.~S.,   {Gilliland} R.~L.,  2003, \mn@doi [\apj] {10.1086/379609}, \href {https://ui.adsabs.harvard.edu/abs/2003ApJ...599.1320K} {599, 1320}

\bibitem[\protect\citeauthoryear{{Leiner}, {Geller}, {Gully-Santiago}, {Gosnell}  \& {Tofflemire}}{{Leiner} et~al.}{2022}]{2022ApJ...927..222L}
{Leiner} E.~M.,  {Geller} A.~M.,  {Gully-Santiago} M.~A.,  {Gosnell} N.~M.,   {Tofflemire} B.~M.,  2022, \mn@doi [\apj] {10.3847/1538-4357/ac53b1}, \href {https://ui.adsabs.harvard.edu/abs/2022ApJ...927..222L} {927, 222}

\bibitem[\protect\citeauthoryear{{Libralato} et~al.,}{{Libralato} et~al.}{2022}]{Libralato22}
{Libralato} M.,  et~al., 2022, \mn@doi [\apj] {10.3847/1538-4357/ac7727}, \href {https://ui.adsabs.harvard.edu/abs/2022ApJ...934..150L} {934, 150}

\bibitem[\protect\citeauthoryear{{Lugger}, {Cohn}, {Heinke}, {Grindlay}  \& {Edmonds}}{{Lugger} et~al.}{2007}]{Lugger07}
{Lugger} P.~M.,  {Cohn} H.~N.,  {Heinke} C.~O.,  {Grindlay} J.~E.,   {Edmonds} P.~D.,  2007, \mn@doi [\apj] {10.1086/507572}, \href {http://adsabs.harvard.edu/abs/2007ApJ...657..286L} {657, 286}

\bibitem[\protect\citeauthoryear{{Lugger}, {Cohn}, {Heinke}, {Zhao}, {Zhao}  \& {Anderson}}{{Lugger} et~al.}{2023}]{Lugger23}
{Lugger} P.~M.,  {Cohn} H.~N.,  {Heinke} C.~O.,  {Zhao} J.,  {Zhao} Y.,   {Anderson} J.,  2023, \mn@doi [\mnras] {10.1093/mnras/stad1887}, \href {https://ui.adsabs.harvard.edu/abs/2023MNRAS.tmp.1798L} {}

\bibitem[\protect\citeauthoryear{{Luna}, {Sokoloski}, {Mukai}  \& {Nelson}}{{Luna} et~al.}{2013}]{2013A&A...559A...6L}
{Luna} G.~J.~M.,  {Sokoloski} J.~L.,  {Mukai} K.,   {Nelson} T.,  2013, \mn@doi [\aap] {10.1051/0004-6361/201220792}, \href {https://ui.adsabs.harvard.edu/abs/2013A&A...559A...6L} {559, A6}

\bibitem[\protect\citeauthoryear{{Margon} et~al.,}{{Margon} et~al.}{2010}]{Margon10}
{Margon} B.,  et~al., 2010, in {Kalogera} V.,  {van der Sluys} M.,  eds,  American Institute of Physics Conference Series Vol. 1314, International Conference on Binaries: in celebration of Ron Webbink's 65th Birthday. pp 163--168 (\mn@eprint {arXiv} {1011.1005}), \mn@doi{10.1063/1.3536360}

\bibitem[\protect\citeauthoryear{{Mathieu}, {van den Berg}, {Torres}, {Latham}, {Verbunt}  \& {Stassun}}{{Mathieu} et~al.}{2003}]{Mathieu03}
{Mathieu} R.~D.,  {van den Berg} M.,  {Torres} G.,  {Latham} D.,  {Verbunt} F.,   {Stassun} K.,  2003, \mn@doi [\aj] {10.1086/344944}, \href {https://ui.adsabs.harvard.edu/abs/2003AJ....125..246M} {125, 246}

\bibitem[\protect\citeauthoryear{{Maxwell}, {Lugger}, {Cohn}, {Heinke}, {Grindlay}, {Budac}, {Drukier}  \& {Bailyn}}{{Maxwell} et~al.}{2012}]{Maxwell12}
{Maxwell} J.~E.,  {Lugger} P.~M.,  {Cohn} H.~N.,  {Heinke} C.~O.,  {Grindlay} J.~E.,  {Budac} S.~A.,  {Drukier} G.~A.,   {Bailyn} C.~D.,  2012, \mn@doi [\apj] {10.1088/0004-637X/756/2/147}, \href {https://ui.adsabs.harvard.edu/abs/2012ApJ...756..147M} {756, 147}

\bibitem[\protect\citeauthoryear{{Nardiello} et~al.,}{{Nardiello} et~al.}{2018}]{Nardiello18}
{Nardiello} D.,  et~al., 2018, \mn@doi [\mnras] {10.1093/mnras/sty2515}, \href {https://ui.adsabs.harvard.edu/abs/2018MNRAS.481.3382N} {481, 3382}

\bibitem[\protect\citeauthoryear{{Oh}, {Hui}, {Li}  \& {Kong}}{{Oh} et~al.}{2020}]{Oh20}
{Oh} K.,  {Hui} C.~Y.,  {Li} K.~L.,   {Kong} A.~K.~H.,  2020, \mn@doi [\mnras] {10.1093/mnras/staa2462}, \href {https://ui.adsabs.harvard.edu/abs/2020MNRAS.498..292O} {498, 292}

\bibitem[\protect\citeauthoryear{{Piotto} et~al.,}{{Piotto} et~al.}{2015a}]{Piotto15}
{Piotto} G.,  et~al., 2015a, \mn@doi [\aj] {10.1088/0004-6256/149/3/91}, \href {https://ui.adsabs.harvard.edu/abs/2015AJ....149...91P} {149, 91}

\bibitem[\protect\citeauthoryear{{Piotto} et~al.,}{{Piotto} et~al.}{2015b}]{2015AJ....149...91P}
{Piotto} G.,  et~al., 2015b, \mn@doi [\aj] {10.1088/0004-6256/149/3/91}, \href {https://ui.adsabs.harvard.edu/abs/2015AJ....149...91P} {149, 91}

\bibitem[\protect\citeauthoryear{{Pooley} \& {Hut}}{{Pooley} \& {Hut}}{2006}]{Pooley06}
{Pooley} D.,  {Hut} P.,  2006, \mn@doi [\apjl] {10.1086/507027}, \href {https://ui.adsabs.harvard.edu/abs/2006ApJ...646L.143P} {646, L143}

\bibitem[\protect\citeauthoryear{{Rozyczka}, {Thompson}, {Narloch}, {Pych}  \& {Schwarzenberg-Czerny}}{{Rozyczka} et~al.}{2016}]{Rozyczka16}
{Rozyczka} M.,  {Thompson} I.~B.,  {Narloch} W.,  {Pych} W.,   {Schwarzenberg-Czerny} A.,  2016, \actaa, \href {https://ui.adsabs.harvard.edu/abs/2016AcA....66..307R} {66, 307}

\bibitem[\protect\citeauthoryear{{Rutledge}, {Bildsten}, {Brown}, {Pavlov}  \& {Zavlin}}{{Rutledge} et~al.}{2002}]{Rutledge02a}
{Rutledge} R.~E.,  {Bildsten} L.,  {Brown} E.~F.,  {Pavlov} G.~G.,   {Zavlin} V.~E.,  2002, \mn@doi [\apj] {10.1086/342306}, \href {https://ui.adsabs.harvard.edu/abs/2002ApJ...578..405R} {578, 405}

\bibitem[\protect\citeauthoryear{{Saito}, {Kawai}, {Kamae}, {Shibata}, {Dotani}  \& {Kulkarni}}{{Saito} et~al.}{1997}]{Saito97}
{Saito} Y.,  {Kawai} N.,  {Kamae} T.,  {Shibata} S.,  {Dotani} T.,   {Kulkarni} S.~R.,  1997, \mn@doi [\apjl] {10.1086/310512}, \href {https://ui.adsabs.harvard.edu/abs/1997ApJ...477L..37S} {477, L37}

\bibitem[\protect\citeauthoryear{{Servillat} et~al.,}{{Servillat} et~al.}{2008}]{Servillat08}
{Servillat} M.,  et~al., 2008, \mn@doi [\aap] {10.1051/0004-6361:200810188}, \href {https://ui.adsabs.harvard.edu/abs/2008A&A...490..641S} {490, 641}

\bibitem[\protect\citeauthoryear{{Shishkovsky} et~al.,}{{Shishkovsky} et~al.}{2018}]{Shishkovsky18}
{Shishkovsky} L.,  et~al., 2018, \mn@doi [\apj] {10.3847/1538-4357/aaadb1}, \href {https://ui.adsabs.harvard.edu/abs/2018ApJ...855...55S} {855, 55}

\bibitem[\protect\citeauthoryear{{Taylor}}{{Taylor}}{2005}]{2005ASPC..347...29T}
{Taylor} M.~B.,  2005, in {Shopbell} P.,  {Britton} M.,   {Ebert} R.,  eds,  Astronomical Society of the Pacific Conference Series Vol. 347, Astronomical Data Analysis Software and Systems XIV. p.~29

\bibitem[\protect\citeauthoryear{{Trager}, {King}  \& {Djorgovski}}{{Trager} et~al.}{1995}]{Trager95}
{Trager} S.~C.,  {King} I.~R.,   {Djorgovski} S.,  1995, \mn@doi [\aj] {10.1086/117268}, \href {https://ui.adsabs.harvard.edu/abs/1995AJ....109..218T} {109, 218}

\bibitem[\protect\citeauthoryear{{Vasiliev} \& {Baumgardt}}{{Vasiliev} \& {Baumgardt}}{2021}]{2021MNRAS.505.5978V}
{Vasiliev} E.,  {Baumgardt} H.,  2021, \mn@doi [\mnras] {10.1093/mnras/stab1475}, \href {https://ui.adsabs.harvard.edu/abs/2021MNRAS.505.5978V} {505, 5978}

\bibitem[\protect\citeauthoryear{{Verbunt}}{{Verbunt}}{2000}]{Verbunt00}
{Verbunt} F.,  2000, in {Pallavicini} R.,  {Micela} G.,   {Sciortino} S.,  eds,  Astronomical Society of the Pacific Conference Series Vol. 198, Stellar Clusters and Associations: Convection, Rotation, and Dynamos. p.~421 (\mn@eprint {arXiv} {astro-ph/9907202}), \mn@doi{10.48550/arXiv.astro-ph/9907202}

\bibitem[\protect\citeauthoryear{{Verbunt} \& {Hut}}{{Verbunt} \& {Hut}}{1987}]{Verbunt87}
{Verbunt} F.,  {Hut} P.,  1987, in {Helfand} D.~J.,  {Huang} J.~H.,  eds,  IAU Symposium 125 Vol. 125, The Origin and Evolution of Neutron Stars. p.~187

\bibitem[\protect\citeauthoryear{{Verbunt}, {Pooley}  \& {Bassa}}{{Verbunt} et~al.}{2008}]{Verbunt08}
{Verbunt} F.,  {Pooley} D.,   {Bassa} C.,  2008, in {Vesperini} E.,  {Giersz} M.,   {Sills} A.,  eds, ~ Vol. 246, Dynamical Evolution of Dense Stellar Systems. pp 301--310 (\mn@eprint {arXiv} {0710.1804}), \mn@doi{10.1017/S1743921308015822}

\bibitem[\protect\citeauthoryear{{Wadiasingh}, {Harding}, {Venter}, {B{\"o}ttcher}  \& {Baring}}{{Wadiasingh} et~al.}{2017}]{Wadiasingh17}
{Wadiasingh} Z.,  {Harding} A.~K.,  {Venter} C.,  {B{\"o}ttcher} M.,   {Baring} M.~G.,  2017, \mn@doi [\apj] {10.3847/1538-4357/aa69bf}, \href {https://ui.adsabs.harvard.edu/abs/2017ApJ...839...80W} {839, 80}

\bibitem[\protect\citeauthoryear{{Wilms}, {Allen}  \& {McCray}}{{Wilms} et~al.}{2000}]{2000ApJ...542..914W}
{Wilms} J.,  {Allen} A.,   {McCray} R.,  2000, \mn@doi [\apj] {10.1086/317016}, \href {https://ui.adsabs.harvard.edu/abs/2000ApJ...542..914W} {542, 914}

\bibitem[\protect\citeauthoryear{{Zavlin}, {Pavlov}, {Sanwal}, {Manchester}, {Tr{\"u}mper}, {Halpern}  \& {Becker}}{{Zavlin} et~al.}{2002}]{Zavlin02}
{Zavlin} V.~E.,  {Pavlov} G.~G.,  {Sanwal} D.,  {Manchester} R.~N.,  {Tr{\"u}mper} J.,  {Halpern} J.~P.,   {Becker} W.,  2002, \mn@doi [\apj] {10.1086/339351}, \href {https://ui.adsabs.harvard.edu/abs/2002ApJ...569..894Z} {569, 894}

\bibitem[\protect\citeauthoryear{{Zhang} et~al.,}{{Zhang} et~al.}{2022}]{Zhang22}
{Zhang} L.,  et~al., 2022, \mn@doi [\apjl] {10.3847/2041-8213/ac81c3}, \href {https://ui.adsabs.harvard.edu/abs/2022ApJ...934L..21Z} {934, L21}

\bibitem[\protect\citeauthoryear{{Zhao} \& {Heinke}}{{Zhao} \& {Heinke}}{2022}]{Zhao22}
{Zhao} J.,  {Heinke} C.~O.,  2022, \mn@doi [\mnras] {10.1093/mnras/stac442}, \href {https://ui.adsabs.harvard.edu/abs/2022MNRAS.511.5964Z} {511, 5964}

\bibitem[\protect\citeauthoryear{{Zhao} et~al.,}{{Zhao} et~al.}{2020}]{Zhao20}
{Zhao} Y.,  et~al., 2020, \mn@doi [\mnras] {10.1093/mnras/staa2927}, \href {https://ui.adsabs.harvard.edu/abs/2020MNRAS.499.3338Z} {499, 3338}

\makeatother
\end{thebibliography}

\section{Appendix A}

\renewcommand{\thefigure}{A\arabic{figure}}

\begin{figure*}
\hspace*{0.090\textwidth}
\makebox[0.425\textwidth][c]{\large $U_{336}$}
\makebox[0.425\textwidth][c]{\large $V_{606}$}
\vcenteredhbox{\begin{minipage}{0.090\textwidth}\large \flushright X1\end{minipage}}
\vcenteredhbox{\includegraphics[width=0.8\textwidth]{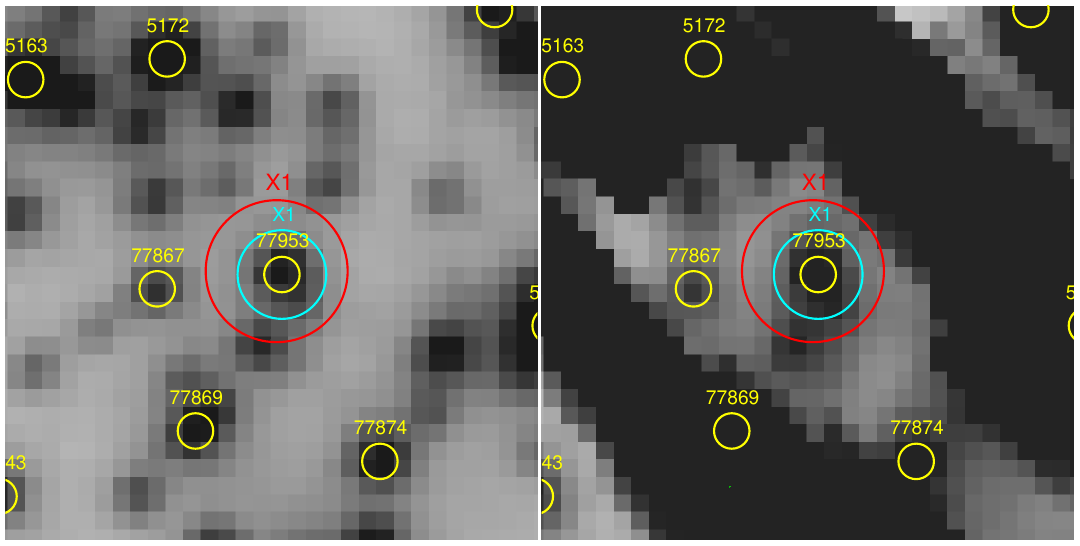}}
\vcenteredhbox{\begin{minipage}{0.090\textwidth}\large \flushright X4\end{minipage}}
\vcenteredhbox{\includegraphics[width=0.8\textwidth]{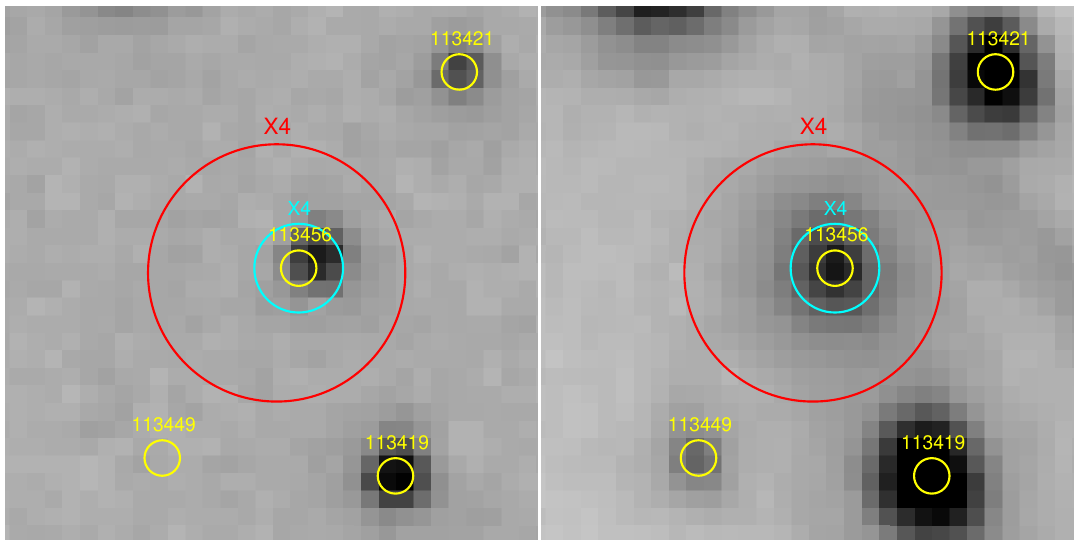}}
\vcenteredhbox{\begin{minipage}{0.090\textwidth}\large \flushright X9\end{minipage}}
\vcenteredhbox{\hspace*{0.025in}\includegraphics[width=0.8\textwidth]{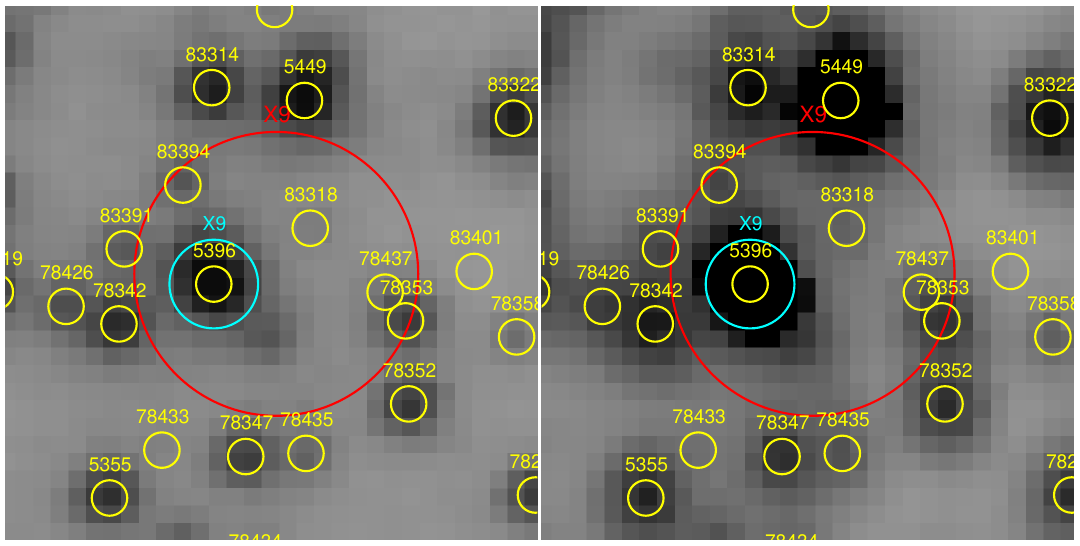}}
\caption{Finding charts: X1, X4, and X9. The field size is 1.2 arcsec on a side. North is up and east is to the left. HUGS stars are indicated by yellow circles labelled by the HUGS sequence number. The red circles are the \emph{Chandra} error circles as computed (see text and Table 3). The cyan circles indicate the locations of the proposed counterparts. }  \label{fig:finding_charts_1}
\end{figure*}

\newpage

\begin{figure*}
\hspace*{0.090\textwidth}
\makebox[0.45\textwidth][c]{\large $U_{336}$}
\makebox[0.45\textwidth][c]{\large $V_{606}$}
\vcenteredhbox{\begin{minipage}{0.090\textwidth}\large \flushright X10\end{minipage}}
\vcenteredhbox{\hspace*{0.025in}\includegraphics[width=0.8\textwidth]{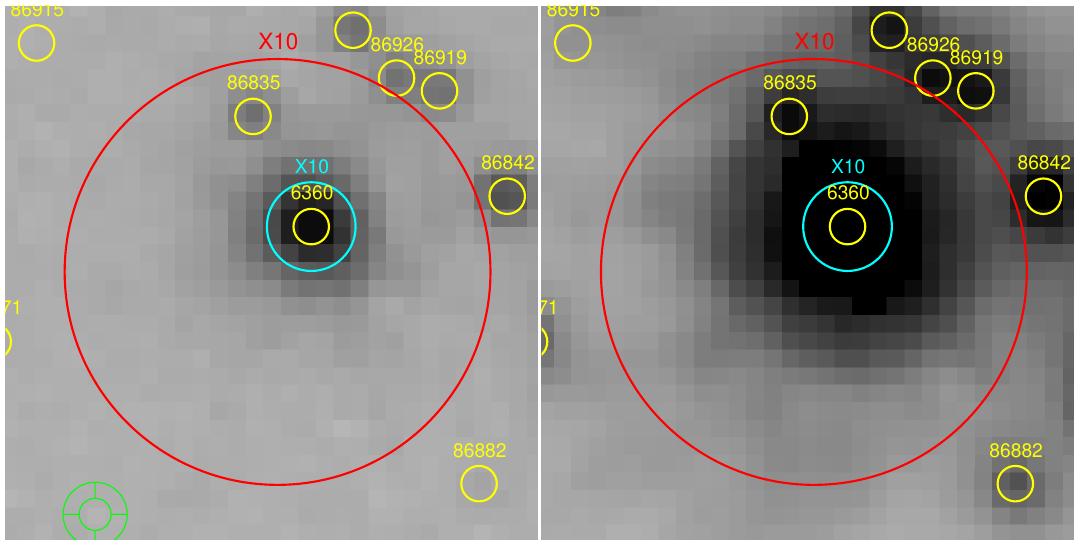}}
\vcenteredhbox{\begin{minipage}{0.090\textwidth}\large \flushright X12\end{minipage}}
\vcenteredhbox{\hspace*{0.025in}\includegraphics[width=0.8\textwidth]{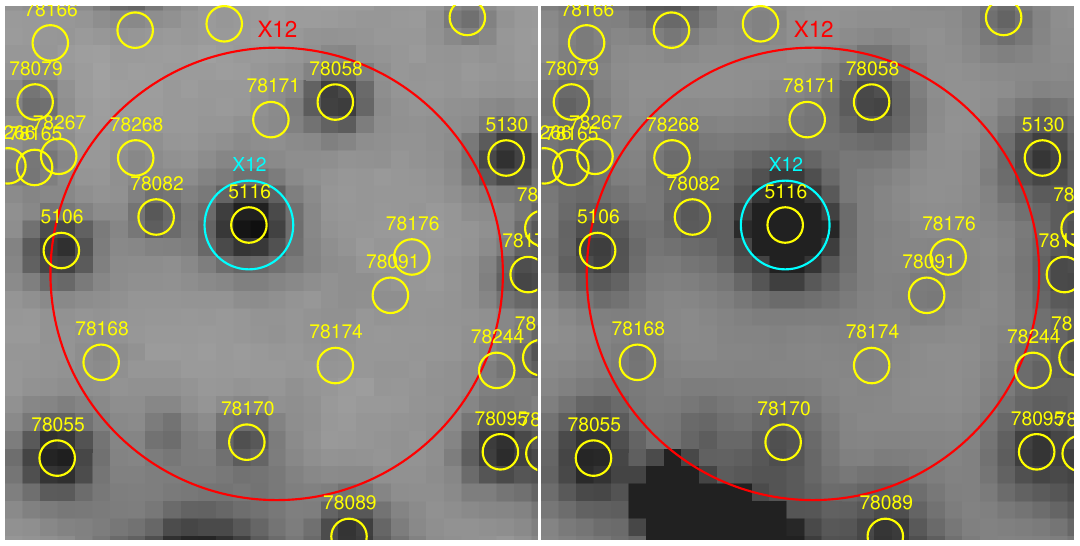}}
\vcenteredhbox{\begin{minipage}{0.090\textwidth}\large \flushright X15\end{minipage}}
\vcenteredhbox{\hspace*{0.025in}\includegraphics[width=0.8\textwidth]{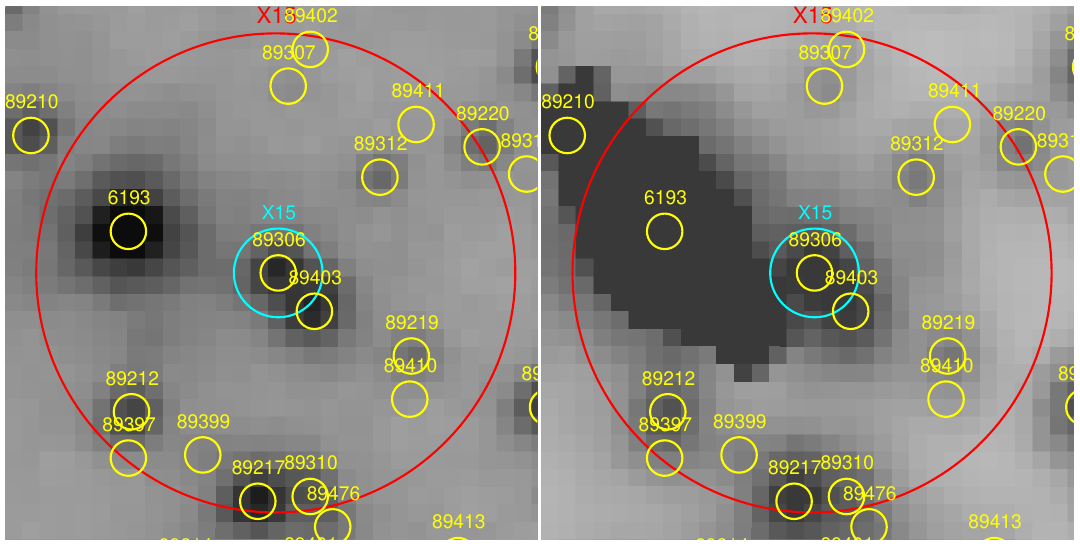}}
\caption{Finding charts: X10, X12, and X15. The field size is 1.2 arcsec on a side. }  \label{fig:finding_charts_2}
\end{figure*}

\newpage

\begin{figure*}
\hspace*{0.090\textwidth}
\makebox[0.45\textwidth][c]{\large $U_{336}$}
\makebox[0.45\textwidth][c]{\large $V_{606}$}
\vcenteredhbox{\begin{minipage}{0.090\textwidth}\large \flushright X16\end{minipage}}
\vcenteredhbox{\includegraphics[width=0.8\textwidth]{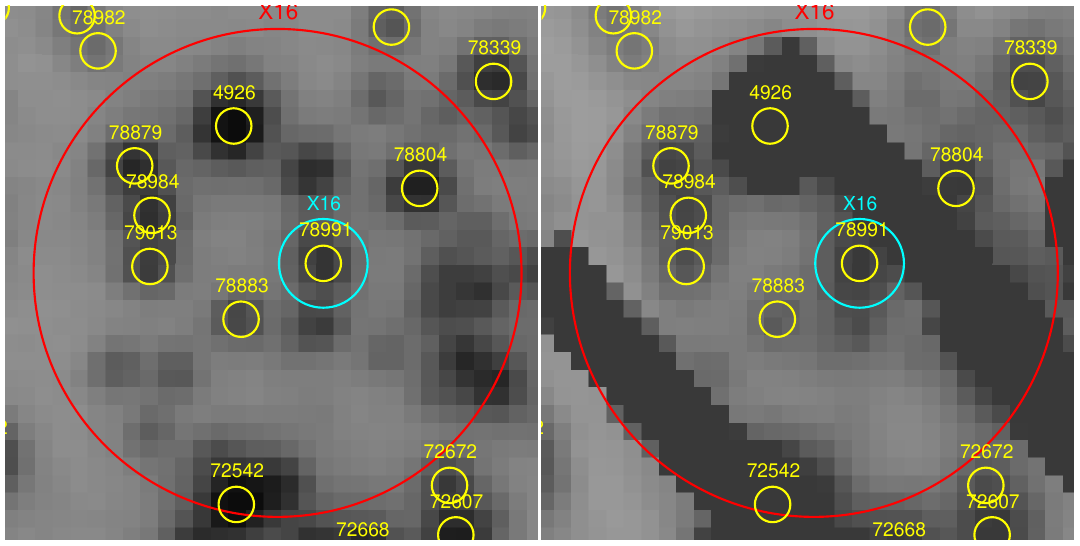}}
\vcenteredhbox{\begin{minipage}{0.090\textwidth}\large \flushright X20\end{minipage}}
\vcenteredhbox{\includegraphics[width=0.8\textwidth]{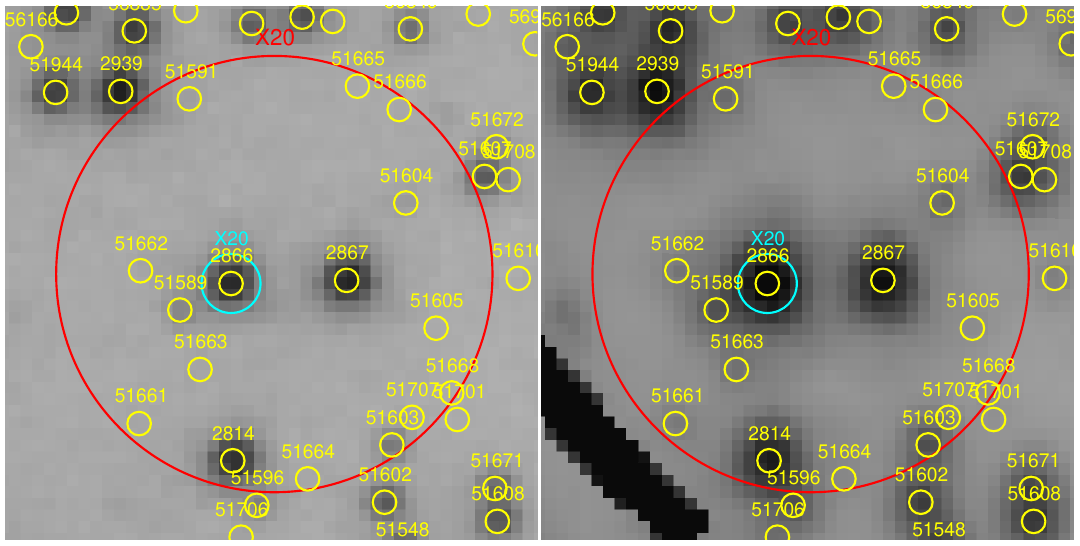}}
\vcenteredhbox{\begin{minipage}{0.090\textwidth}\large \flushright X21\end{minipage}}
\vcenteredhbox{\hspace*{0.025in}\includegraphics[width=0.8\textwidth]{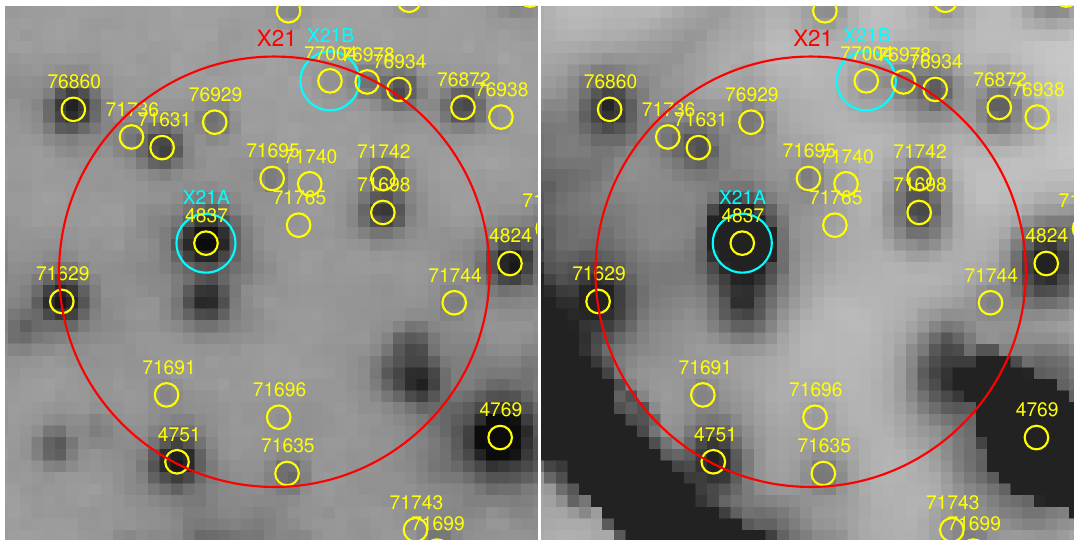}}
\caption{Finding charts: X16, X20, and X21. The field size is 1.2 arcsec on a side for X16, and 1.8 arcsec on a side for X20 and X21. }  \label{fig:finding_charts_3}
\end{figure*}

\newpage

\begin{figure*}
\hspace*{0.090\textwidth}
\makebox[0.45\textwidth][c]{\large $U_{336}$}
\makebox[0.45\textwidth][c]{\large $V_{606}$}
\vcenteredhbox{\begin{minipage}{0.090\textwidth}\large \flushright X23\end{minipage}}
\vcenteredhbox{\includegraphics[width=0.8\textwidth]{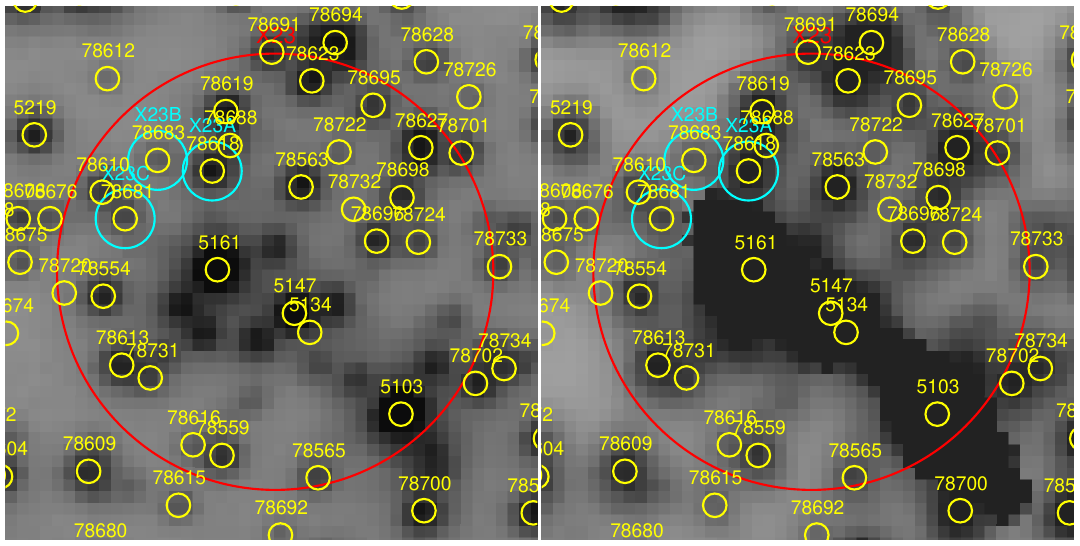}}
\vcenteredhbox{\begin{minipage}{0.090\textwidth}\large \flushright X24\end{minipage}}
\vcenteredhbox{\includegraphics[width=0.8\textwidth]{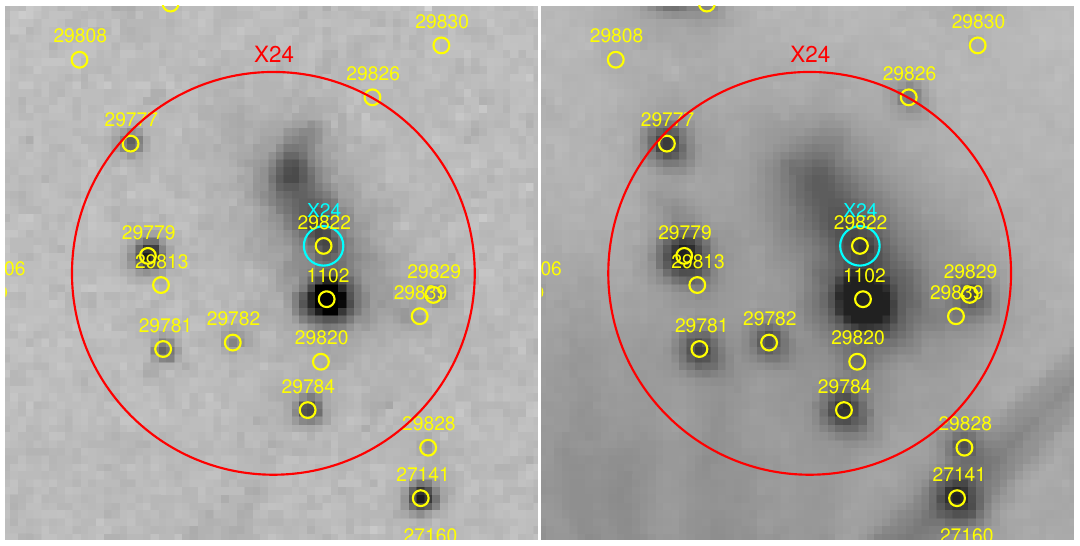}}
\vcenteredhbox{\begin{minipage}{0.090\textwidth}\large \flushright X26\end{minipage}}
\vcenteredhbox{\hspace*{0.025in}\includegraphics[width=0.8\textwidth]{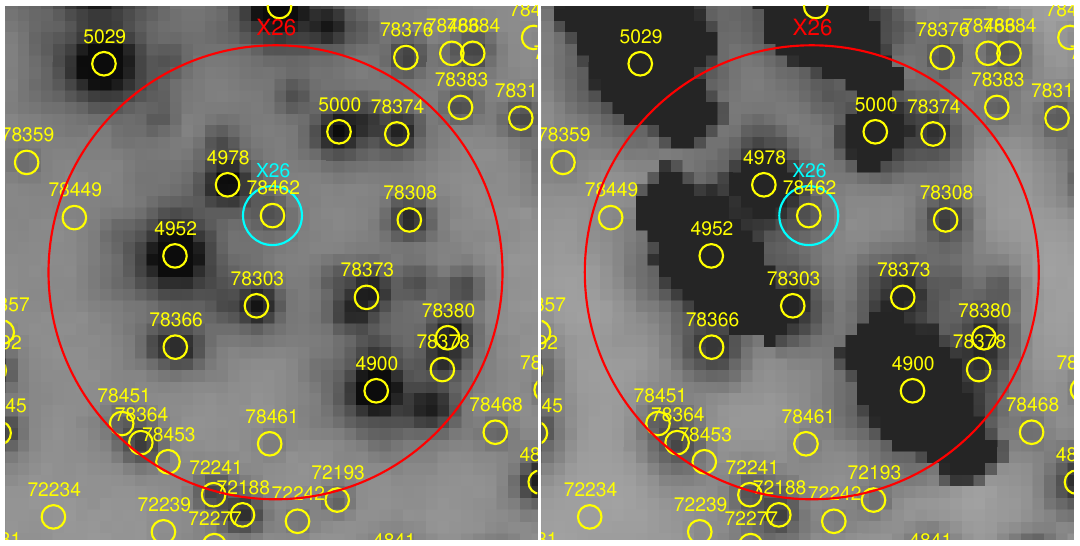}}
\caption{Finding charts: X23. X24, and X26. The field size is 1.8 arcsec on a side for X23 and X26, and 2.7 arcsec on a side for X24. } \label{fig:finding_charts_4}
\end{figure*}

                                                 %
                                                 %
\bsp	
\label{lastpage}                                 %
\end{document}